\documentclass[]{aastex631}

\definecolor{Purple}{rgb}{1,0,1}

\usepackage{longtable}
\usepackage{rotating}

\begin{document}

\title{What Causes Errors in Wang-Sheeley-Arge Solar Wind Modeling at L1 ?}

\author[0000-0002-6553-3807]{Satabdwa Majumdar}
\affiliation{Austrian Space Weather Office, GeoSphere Austria, Graz, 8020, Austria}

\author[0000-0002-6362-5054]{Martin A. Reiss}
\affiliation{Community Coordinated Modeling Center, NASA Goddard Space Flight Center, 8800 Greenbelt Rd., Greenbelt, MD 20771, USA}

\author[0000-0002-5547-9683]{Karin Muglach}
\affiliation{NASA Goddard Space Flight Center, 8800 Greenbelt Rd., Greenbelt, MD 20771, USA}
\affiliation{Catholic University of America, Washington, DC 20064, USA}

\author[0000-0001-9326-3448]{Charles N. Arge}
\affiliation{NASA Goddard Space Flight Center, 8800 Greenbelt Rd., Greenbelt, MD 20771, USA}



\begin{abstract}
Previous ambient solar wind (SW) validation studies have reported on discrepancies between modeled and observed SW conditions at L1. They indicated that a major source of discrepancies stems from how we model the solar corona. Thus, enhancing predictive capabilities demands a thorough examination of coronal modeling. The Wang-Sheeley-Arge (WSA) model has been a workhorse model that provides the near-Sun SW conditions. An important component of it is the Potential Field Source Surface (PFSS) model. This study analyzes 15 different Carrington Rotations (CRs), and presents detailed analysis of CR 2052 to identify WSA model settings that lead to successful and erroneous SW predictions at Earth. For the events studied, we show that increasing the model’s grid resolution improves the open-close boundary identification. This results in better predicting the onset and duration of high-speed streams (HSSs). In addition, we find an optimized source surface height (R$_{ss}$) (lying between 1.8-3.1 R$_{\odot}$) further enhances HSS prediction accuracy for the studied events. A detailed analysis shows that changes in R$_{ss}$, (a) changes the Great Circle Angular Distance (GCAD) maps (at the solar surface) of the associated coronal holes and (b) changes the foot-point locations of the magnetic connectivities to the sub-Earth locations. These factors change the near-Sun SW speed, that eventually leads to uncertainties in speeds near Earth. We also investigate the usefulness of coronal hole observations in constraining R$_{ss}$ and SW solutions at Earth, and highlight their underutilized value in guiding the selection of magnetic maps for improved ambient solar wind modeling at L1. 
\end{abstract}


\section{Introduction}\label{intro}

The solar wind consists of a stream of plasma embedded in a magnetic field which originates at the base of the hot solar corona that continuously propagates outward and forms the heliosphere. This stream of plasma and magnetic field shows a wide range of physical properties in terms of their speed, density, temperature, strength and orientation of the embedded magnetic field and exhibits a variety of physical processes \citep[for reviews, see][]{Cranmer2017SS,Verscharen2019LRSP}. Predominantly, the solar wind has a bi-modal nature in terms of their speed, having a slow and a fast component as measured at L1~\citep[e.g.][]{Larrodera2020A&A}. Out of these two components, it is the latter one, which is called a High Speed Stream (HSS) which is important for space weather, as fast solar wind streams contribute to about 70\% of geomagnetic activity outside the solar maximum and about 30\% during solar maximum~\citep{Richardson2000JGR}. Traveling through this ambient solar wind, coronal Mass Ejections (CMEs) are the most dramatic source of space weather. After being ejected at the Sun's corona, CMEs propagate through the ambient (fast or slow) solar wind and experience a drag force originating from the interaction with the solar wind that can lead to an acceleration or deceleration in CMEs, modulating their arrival times at Earth and other planetary objects~\citep[see e.g.~][]{Gopalswamy2000GeoRL,Majumdar2020ApJ,Majumdar2021ApJ}. Apart from CMEs, the ambient solar wind (with the embedded magnetic field) also defines the path along which solar energetic particles (SEPs) which are also major sources of ground-level enhancements \citep{Reames2023FrASS}. All these factors have made the knowledge of the evolving solar wind indispensable for successful space weather research and forecasting \citep{Owens2013LRSP}.

A connection between enhanced geomagnetic activity and open coronal magnetic fields was first suggested by \cite{Billings1964ApNr}, which was soon followed by \cite{Altschuler1972SoPh}, who modeled the solar coronal field and reported that open diverging unipolar coronal magnetic field lines were connected to coronal holes (identified as regions of lower density and temperature compared to the surrounding coronal plasma). \cite{Krieger1973SoPh} further reported that some of these coronal holes are associated with an HSS. Taken together, these studies established that coronal holes, which are characterized by open magnetic field topologies, serve as the source of HSSs and pointed out the crucial role of magnetic modeling of the corona in solar wind studies. 

Therefore scientists have developed modeling techniques that see the Sun-Earth system as a series of well connected and coupled parts consisting of the solar photosphere, the corona and the inner heliosphere. Such a coupled modeling system can basically be considered to have a coronal domain and a heliospheric domain, sometimes with an interface region in between the two domains \citep[for an overview, see Figure 1 in][]{Reiss2023AdSpR}. Synoptic maps of the photospheric magnetic field serve as the basic observational input for the coronal domain to specify the inner boundary condition of the coronal model. Synoptic maps from the Global Oscillations Network Group (GONG) produced by the National Solar Observatory (NSO) are the most commonly used input magnetic field maps for solar wind modeling and forecasting. Instead of these standard synoptic maps, ideally it would be best to use synchronic magnetograms which provide a global snapshot of the Sun’s magnetic field at any specific moment in time, but unfortunately such maps are not available with current solar instrumentation. The magnetic maps developed with the Air Force Data Assimilative Photospheric flux Transport (ADAPT) model \citep{Arge2010AIPC} are an attempt to create quasi-synchronic maps which use localized ensemble filtering techniques to adjust a set of photospheric flow simulations (accounting for differential rotation, meridional flows, super granular diffusion), making them agree with available observations \citep[for details on ADAPT maps, please see; ][]{Arge2010AIPC,Hickmann2015SoPh}. 

Owing to the absence of routine measurements of the coronal magnetic field, the coronal domain relies on extrapolation techniques to reproduce the large-scale magnetic field of the solar corona, using the line-of-sight component of the photospheric magnetic field as input. The most widely applied extrapolation technique in the coronal domain is the Potential Field Source Surface (PFSS) model \citep{Altschuler1969SoPh}, which approximates a potential field (or current-free) configuration of the solar corona up to a certain height, called the source surface height. By comparing the modeled magnetic structures with coronagraph observations, the best agreement was found for a source surface height of 2.5~R$_{\odot}$ \citep{Altschuler1969SoPh}. Advanced models in the coronal domain account for the dynamic behavior of the solar wind plasma by solving a set of non-linear magnetohydrodynamics (MHD) equations. Notable examples of well-established MHD codes for the coronal domain include the Magnetohydrodynamic Algorithm outside a Sphere (MAS) model \citep{Linker1999JGR,Riley2001JGR}, the Space Weather Modeling Framework (SWMF) \citep{Toth2005JGRA}, CORHEL global coronal and heliospheric modeling suite \citep{Riley2012JASTP}, EUropean Heliospheric FORecasting Information Asset \citep[EUHFORIA; ][]{Pomoell2018JSWSC} and GAMERA \citep{Zhang2019ApJS}. The interface region between the coronal and heliospheric model domains serves as a critical boundary, defining the outer boundary conditions for the coronal domain and the inner boundary conditions for the heliospheric domain. 

If this boundary condition is not provided by the coronal model itself, empirical relationships that are commonly used to specify solar wind speed at this boundary include the Wang-Sheeley (WS) model \citep{Wang1990ApJ}, the Distance from the Coronal Hole Boundary (DCHB) model \citep{Riley2001JGR}, and the Wang-Sheeley-Arge (WSA) model \citep{Arge2003AIPC}. The traditional WS model is based on an inverse relationship between the flux tube expansion rate of open magnetic field lines and the solar wind speed measured at Earth. The DCHB model, on the other hand, correlates the solar wind speed at the solar surface with the great-circle angular distance to the nearest coronal hole boundary. The WSA model combines both the expansion factor of the magnetic field topology and the minimum distance to the nearest coronal hole boundary, offering a more comprehensive approach. In the heliospheric domain, based on the techniques used and on the increasing level of complexity, there are models of kinematic mapping \citep[see: ][]{Arge2000JGR}, one dimensional upwind extrapolations and global heliospheric MHD modeling \citep[see Enlil, MAS, SWMF, EUHFORIA; ][]{Linker1999JGR, Riley2001JGR,Odstrcil2003AdSpR,Toth2005JGRA,Pomoell2018JSWSC}. The one dimensional upwind extrapolation technique bridges the gap between kinematic mapping and global MHD modeling by simplifying the fluid momentum equation to its most essential components. By neglecting the pressure gradient and gravitational terms, it allows solar wind flows to evolve and interact in a self-consistent manner, capturing the key dynamics while reducing computational complexity \citep[see HUX, THUX, HUXt; ][]{Riley2011SoPh,Reiss2020ApJ,Barnard2022FrP}. For a discussion on the different solar wind models for the heliosphere see \cite{MacNeice2018SpWea}.

In the last decades, advances in observational and modeling capabilities have furthered our understanding of coronal holes and the associated HSSs \citep{Cranmer2009LRSP}. However, the predictions of HSSs are reported to be only slightly better than the recurrence model, a benchmark model that assumes that the solar wind conditions repeat after each solar rotation \citep{Riley2015SpWea}. Validation studies evaluating the performance of state-of-the-art solar wind models reveal typical errors in predicting the arrival times of HSSs of approximately one to two days \citep{Owens2008SpWea,Jian2011SoPh}. This corresponds to about 25–50\% of the travel time of the solar wind from the Sun to Earth \citep{Reiss2016SpWea}. Recognizing the need for improved accuracy, the recent space weather road map emphasized the importance of understanding the space weather origins at the Sun and modeling the coronal magnetic field \citep{Schrijver2015AdSpR}. Since the dynamic pressure term in the fluid momentum equation ($\sim\,\rho v^2$) largely dictates the structure of the solar wind, errors in the near-Sun speed map (produced by the amalgamation of coronal and interface modeling domains) will have a significant impact on the accuracy of the heliospheric solutions \citep{Gonzi2021SpWea}. Therefore, identifying the sources of errors originating in the coronal domain and the interface domain becomes of pivotal importance. Addressing this crucial aspect, studies in the past have shown that possible sources of error could include the choice of the input magnetic maps \citep{Riley2012JASTP,Gressl2014SoPh,Riley2014SoPh,Li2021JGRA,Jin2022SpWea,Demidov2023SoPh,Perri2023ApJ,wang2024ApJ}, pre-processing of magnetic maps \citep{toth2011ApJ} and the choice of the coronal models themselves \citep{toth2011ApJ,Gressl2014SoPh,Jian2015SpWea,MacNeice2018SpWea,Jones2024ApJ,Asvestari2024ApJ}. 

The WSA model has been one of the workhorse models in the community which is used not only for science studies, but also at forecasting centers like National Oceanic and Atmospheric Administration's (NOAA's) Space Weather Prediction Center (SWPC), and the Met Office Space Weather Operations Center. A crucial element of the WSA model is the PFSS model that models the large scale coronal magnetic field. Thus, understanding the effect of this model on solar wind modeling is important. A key aspect of the PFSS model is the source surface height $R_{ss}$. \cite{Altschuler1969SoPh} found that by comparing the PFSS modeled coronal magnetic structures with coronagraph observations, the best agreement was achieved for a source surface height of $R_{ss}$ = 2.5 R$_{\odot}$, which is generally considered the standard value for the source surface height. However, it was recently proposed that the source surface height does not necessarily remain constant but seemingly ``breathes" in and out over the solar cycle \citep{Arden2014JGRA}. Varying the height of the source surface has also led to better agreement between model and observations in terms of the interplanetary magnetic field strength, coronal holes, open magnetic flux and magnetic polarities \citep{Lee2011SoPh,Arden2014JGRA,Asvestari2019JGRA,Nikolic2019SpWea,Badman2020ApJS}.
\cite{Koukras2025A&A} found that the source surface height is responsible for producing the largest uncertainties in locating the source regions of HSSs. \cite{Meadors2020SpWea} used data assimilation with particle filtering to show that adjusting the height of the source surface can improve solar wind predictions. These studies indicate the importance of the source surface height in solar wind modeling, however a clear understanding of the reason for such improvement is yet to be achieved. Thus, to improve our solar wind prediction capabilities, a clear understanding of the different sources of errors and uncertainties arising from the PFSS-WSA model is essential. The current study sheds light on this particular aspect. It offers a detailed analysis of the underlying physics involved in understanding the effect of a varying source surface height for modeling the arrival of an HSS at Earth. The current study also explores the role of the model's grid resolution in determining the accuracy of HSS arrival. We also explore the importance of the choice of the date of creation of ADAPT maps in the context of accurate solar wind modeling. As a possible way forward, we also discuss the usefulness of incorporating coronal hole observations to minimize the sources of uncertainties and constrain the model predictions.

In Section~\ref{method-data} we outline the data sources that we have used in this study, the modeling framework and the working method, followed by the results of this study in Section~\ref{results}, and the main conclusions from our analysis in Section~\ref{conclusion}.

\section{Methodology}\label{method-data}
\subsection{Data}
We use full disk GONG/NSO photospheric magnetic field maps as input to our model. These magnetic maps are measured in Gauss, and are given in a $sin(\theta)-\phi$ grid, with 180 $\times$ 360 grid points for latitude and longitude, respectively. The GONG magnetic maps are available at their online platform\footnote{https://gong.nso.edu/}, either as near real time maps or full Carrington Rotation (CR) maps. In the first part of our study we use the CR maps. In the second part of our study, we also use GONG magnetograms devolved with the ADAPT model \citep{Arge2010AIPC}. Observational data of coronal holes (CHs), are provided by the Extreme ultraviolet Imaging Telescope \citep[EIT; ][]{Delaboudini1995SoPh} onboard the SOlar and Heliospheric Observatory \citep[SOHO; ][]{Domingo1995SoPh}, and the Atmospheric Imaging Assembly \citep[AIA; ][]{Lemen2012SoPh} onboard the Solar Dynamics Observatory \citep[SDO; ][]{Pesnell2012SoPh}. Instead of full-disk Extreme UltraViolet (EUV) images, EUV synoptic maps make it easier to compare coronal holes obtained from model solutions to the observed coronal holes. Therefore, we use synoptic EUV maps based on SOHO/EIT and SDO/AIA full disk images that are available online\footnote{http://satdat.oulu.fi/solar\_data/}. These synoptic maps have a resolution of 3600 $\times$ 1440 pixels, and are available in both linear and sin(latitude) scales \citep[for details on the construction of these synoptic EUV maps, please see; ][]{Hamada2020SoPh}.  For finally validating our solar wind model results with in-situ measurements, we use the hourly-averaged observed solar wind speed data from the OMNIWeb\footnote{https://omniweb.gsfc.nasa.gov/ow.html} database \citep{Papitashvili2014}.

\subsection{Model} \label{sec-model}
The entire modeling domain consists of three components, the coronal domain, a model interface which is followed by the heliospheric domain. The modeling framework begins with an input magnetic map for which we use GONG CR maps that have a grid size of 180$^{\circ}$  $\times$ 360$^{\circ}$  in latitude and longitude. In operational space weather forecast models, particularly for solar wind models, GONG synoptic maps are one amongst the standard choice of input maps, and to make our analysis relevant to such operational models, we also use GONG synoptic maps. For the coronal modeling domain, we use the PFSS model \citep{Altschuler1969SoPh}, in particular the python module \textit{pfsspy} \citep{Stansby2020}. This model uses a finite difference method \citep[based on ][]{van2000ApJ} to calculate the magnetic vector potential in a grid uniformly spaced in ln(r), $\mathrm{cos(\theta),\, \phi}$ in the spherical coordinates (r, $\mathrm{\theta},\,\phi$). The grid sizes along the three directions can be adjusted as needed. The field lines between 1R$_{\odot}$ and the source surface radius (R$_{ss}$) are traced in two ways. We trace the field lines from the inner boundary (at 1R$_{\odot}$) to the outer boundary (at R$_{ss}$) where both open and closed field lines are traced to obtain the modeled coronal hole. We also trace the open field lines from the outer boundary to the inner boundary, where the field lines end up inside the boundaries of the modeled coronal holes. The output from the PFSS model is then coupled at the interface region with the Wang-Sheeley-Arge (WSA) empirical relation \citep{Arge2003AIPC} which provides solar wind conditions near the Sun at the outer boundary of the PFSS model (which is the source surface height). This empirical relation relies on unifying two important aspects of the coronal magnetic field to derive the solar wind speed, which are the magnetic fluxtube expansion factor (EF) \citep{Wang1990ApJ}, and the great circle angular distance (GCAD) from the nearest open-closed boundary.

The fluxtube EF indicates by how much a magnetic fluxtube has expanded between the photosphere and a reference height (in our case R$_{ss}$), and is mathematically
expressed as:

\begin{equation}\label{eqn-ef}
    f_p = \left(\frac{R_{\odot}}{R_{ss}}\right)^2 \left|\frac{B(R_{\odot},\theta_{\odot},\phi_{\odot})}{B(R_{ss},\theta_{ss},\phi_{ss})}\right|
\end{equation}

where, $R_{\odot}$ and $R_{ss}$ denote the height of the photosphere and height of the source surface, and $B(R_{\odot},\theta_{\odot},\phi_{\odot})$ and $B(R_{ss},\theta_{ss},\phi_{ss})$ denote the magnetic field strength at the photosphere and the source surface height, respectively.

The GCAD denote the distance (say $d$) of the foot-points of open magnetic field lines at the Sun from the nearest modeled coronal hole boundary \citep{Riley2001JGR}. Both these quantities are calculated from the field line tracing of the PFSS model. Taking these two factors into account, the WSA relation unifies these two magnetic features to provide the solar wind speed through the following relation,

\begin{equation}\label{eqn-v_wsa}
    v_{wsa}(f_p,d) = c_1 + \frac{c_2}{(1+f_p)^{c_3}} \biggl\{c_4 - c_5exp \biggl[-\biggl(\frac{d}{c_6}\biggr)^{c_7}\biggr] \biggr\}^{c_8}
\end{equation}

where $c_1$ to $c_8$ are model coefficients, which are taken from \cite{Reiss2019ApJS,Reiss2020ApJ} as $c_1 = 285$, $c_2 = 625$, $c_3 = 2/9$, $c_4 = 1$, $c_5 = 0.8$, $c_6 = 2$, $c_7 = 2$ and $c_8 = 3$. Using these coefficients, the near Sun solar wind speed computed from the WSA relation (Equation~\ref{eqn-v_wsa}) is then fed into the heliospheric modeling domain to propagate the large scale solar wind solutions from the source surface out to 1 AU.

For the heliospheric domain we use the Heliospheric Upwind eXtrapolation (HUX) model \citep{Riley2011SoPh}, which is a simple yet effective model that uses a 1D upwind extrapolation technique, simplifying the fluid momentum equation by neglecting the pressure gradient, gravity and the magnetic field terms. In the final step, we compare our modeled solar wind results with in-situ solar wind data from the OMNI database. Some key characteristics of an HSS are the starting time of the HSS ($t_{start}$), the duration of the HSS ($t_{dur}$) which we roughly define as the interval between the start and end time of the HSS, and the maximum speed attained by the HSS ($v_{peak}$). A good prediction is one that effectively addresses these aspects, and minimizes the difference between modeled and observed values in these quantities. In the following section, we discuss different factors that can introduce errors in these key aspects of a modeled HSS and in addition we propose strategies to mitigate these errors and produce more accurate solar wind predictions. 

\section{Results} \label{results}
\subsection{Effect of Grid Resolution}
\begin{figure}
    \centering
    \includegraphics[width=0.8\linewidth]{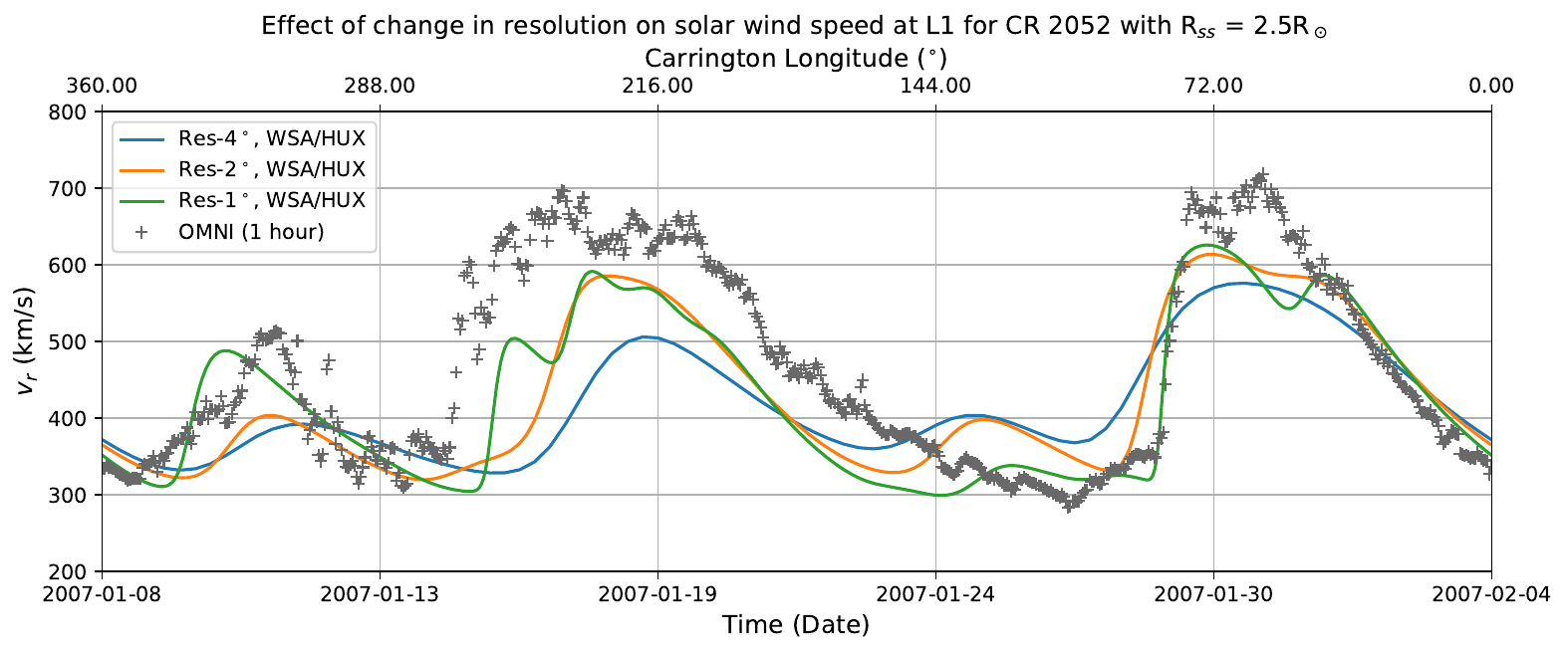}
    \caption{Solar wind speed at L1 for CR 2052 for a WSA/HUX grid resolution of $4^{\circ}$ (in blue), $2^{\circ}$ (in orange) and $1^{\circ}$ (in green) are shown along with the hourly averaged observed speed from the OMNI database (grey crosses).}
    \label{fig-res-vr}
\end{figure}

\begin{figure}
    \centering
    \includegraphics[width=0.6\linewidth]{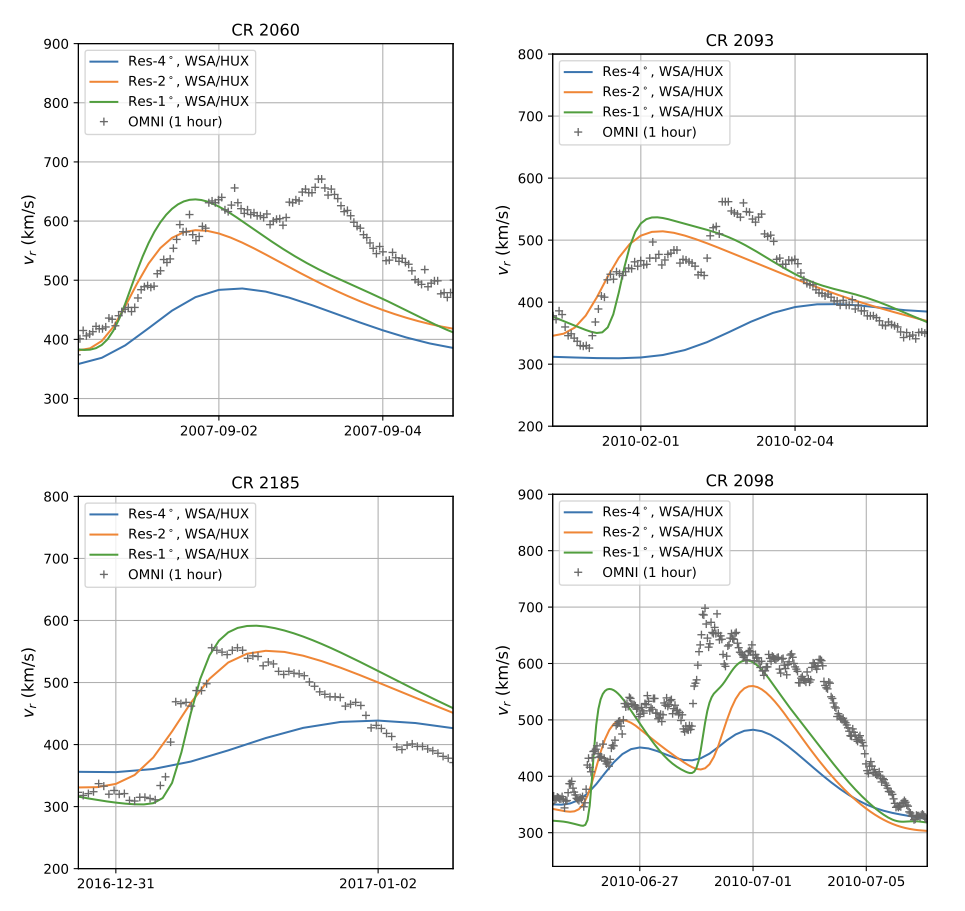}
    \caption{Effect of change in PFSS grid resolution on the modelled solar wind speed profile at L1 for various HSS that were observed during different CRs. The color codes are the same as in Figure~\ref{fig-res-vr}.}
    \label{fig-tstart-stats}
\end{figure}

In Figure~\ref{fig-res-vr} we plot the hourly-averaged observed solar wind speed at L1 from the OMNI database (in grey crosses) for CR 2052. Since every CR starts with Carrington longitude of $360^{\circ}$ and ends with $0^{\circ}$, we plot the dates along the bottom x axis, and the corresponding Carrington longitudes along the top x axis, essentially illustrating that time progresses from left to right in the plot. For this CR, we find two very distinct HSSs being observed at L1 (starting at around January 15 and January 29). We use the GONG CR map for CR 2052 and set the parameters of the PFSS model to a grid resolution of $4^{\circ}$ along both latitude and longitude, and a standard source surface height of
R$_{ss}=\,$2.5R$_{\odot}$. The WSA/HUX model output for these parameters is plotted (in blue). Despite the model successfully predicting both HSSs, noticeable discrepancies can be seen in all three key aspects of the observed and modeled HSS ($t_{start}$, $t_{dur}$ and $v_{peak}$ as mentioned in Section~\ref{sec-model} ). The PFSS model offers the flexibility to change the grid resolution, thus we run the model again with a higher grid resolution of $2^{\circ}$ and $1^{\circ}$ (in both the latitude and longitude). The model output for these resolutions are shown in Figure~\ref{fig-res-vr} (in the orange and green line, respectively). We note a distinct improvement in the modeled HSS: for both HSSs, there is a significant improvement in the $t_{start}$, and hence in $t_{dur}$. For the first HSS, we find that with a grid resolution of $4^{\circ}$, the model was predicting an arrival time that was much later than the actual arrival, while for the second HSS, it was the opposite, thus leading to an under-estimate and an over-estimate in the $t_{dur}$ for the first and the second HSS respectively. Figure~\ref{fig-res-vr} shows that with better resolution, the model is predicting an earlier arrival of the first HSS and a later time for the second, thus getting closer to the observed $t_{start}$ and $t_{dur}$. By improving the grid resolution, improvement in the modeled $t_{start}$ and $t_{dur}$ is achieved, and this improvement is not monotonic, that is to say, an improvement in the resolution is shifting the modeled $t_{start}$ along opposite directions for the two HSSs, bringing them closer to the observed time and thereby significantly reducing the error in $t_{dur}$. In Figure~\ref{fig-tstart-stats} we consider additional CRs which we can connect to an HSS at L1. These CRs were selected at different phases of the solar cycle (including cycle minimum and maximum periods) based on the presence of one or two HSSs that were recorded at L1. Since this study is concerned with the key aspects of HSSs, this figure only includes the period of the HSS (and not the full CR.) The modeled speed profiles for these CRs, for the three different grid resolutions are shown in each case. We again find similar results that improving the resolution leads to improvement in the modeled $t_{start}$ and $t_{dur}$.

\begin{figure}
    \centering
    \includegraphics[width=0.7\linewidth]{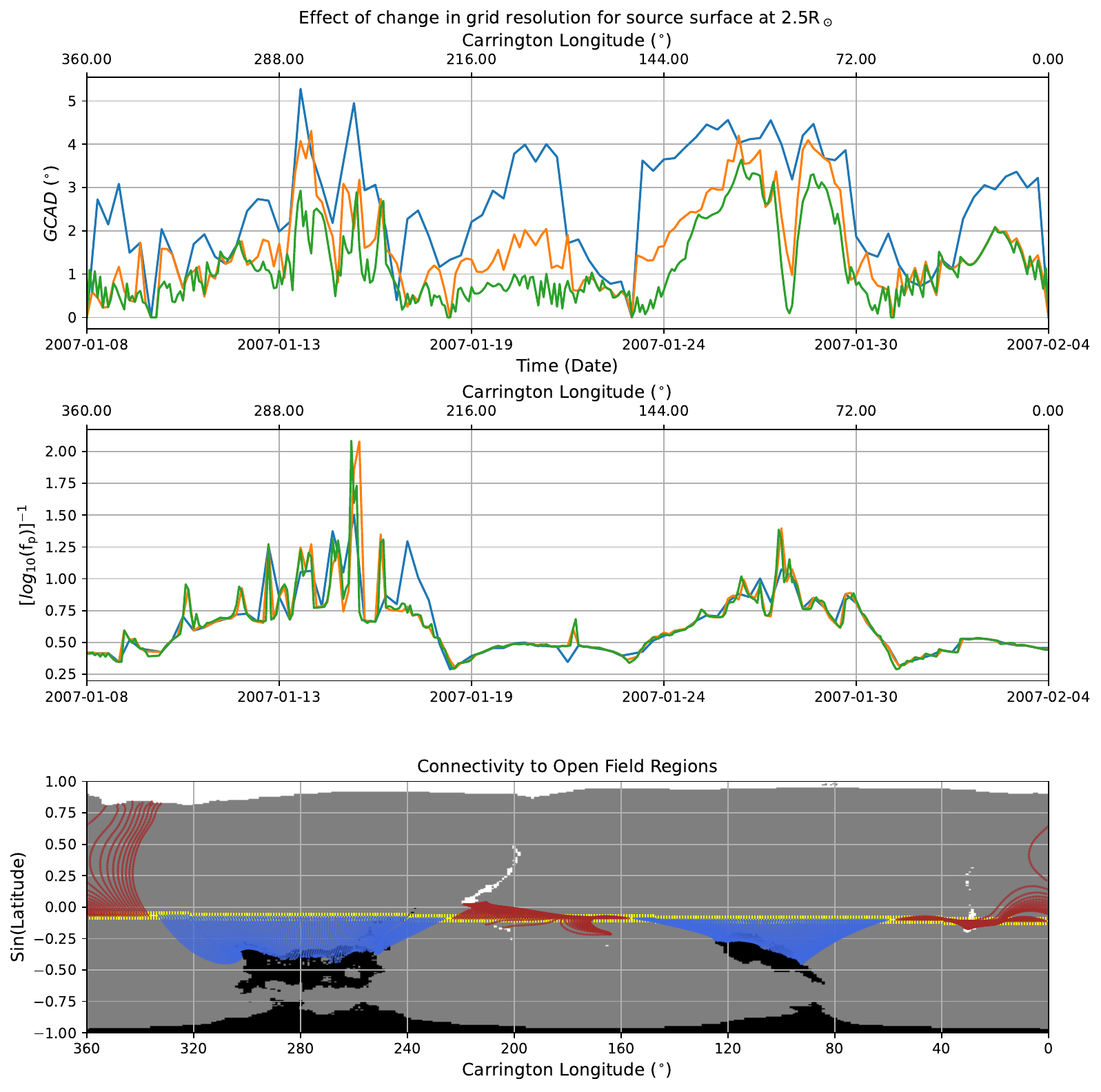}
    \caption{Effect of change in WSA/HUX grid resolution for CR 2052, on the GCAD profile (top panel) and inverse of the logarithm of the expansion factor profile (middle panel) at sub-Earth locations at R$_{ss}$. The bottom panel shows the modeled open field regions for the same CR for a grid resolution of $1^{\circ}$. The yellow crosses denote the sub-Earth locations for the entire CR. The grey regions indicate closed magnetic topology, while the black and white regions indicate open field regions with opposite polarities. The solid red and blue lines indicate open field line connectivities from the sub-Earth locations at the outer boundary of the coronal model (R$_{ss}$ = $2.5$~R$_{\odot}$) to the open field regions at the Sun.}
    \label{fig-res-ef-dmap}
\end{figure}

Additional details of the effect of grid resolution is given in Figure~\ref{fig-res-ef-dmap}. It shows a plot of the GCAD profile (top panel), the inverse of the logarithm of the EF profile (middle panel) at the outer boundary of the coronal model (at R$_{ss}$) at the sub-Earth location and the modeled coronal hole map (bottom panel) for CR 2052 for the 3 grid resolutions. The grey regions indicate closed field regions while the regions in white and black indicate open field regions of opposite polarities. The red and blue lines show the magnetic field line connectivities of the open field regions at the Sun to the sub-Earth locations (shown in yellow crosses) at the source surface. We first note that both the GCAD and the EF profiles show peaks at the two large coronal holes located at Carrington Longitudes of $280^{\circ}$ and $100^{\circ}$. This shows that these two coronal holes are associated with smaller EFs and larger GCAD values, which means these two holes are the sources of the fast solar wind. We would like to point out that in Equation~\ref{eqn-v_wsa}, since $d$ changes with change in the grid resolution, the coefficient $c_6$, which is the normalizing factor for $d$ should also change in order to properly normalize $d$ for the different resolutions. In other words, if the same value of $c_6$ is used for different resolutions, the GCAD values will not be normalized properly. For example, if the $c_6$ value that is used for a lower resolution is also used for a higher resolution, it would lead to an over-normalizing of the $d$ parameter in Equation~\ref{eqn-v_wsa} which would flatten the GCAD profile. Recently, addressing this issue, \cite{Mayank2022ApJS} suggested a generalized approach of using the median of $d$ as $c_6$ for each resolution, where the median value of $d$ is calculated by considering only the open field lines that reach the sub-Earth locations for a particular CR. We follow the same approach in our work. Giving a closer look to the top two panels, we find that changing the grid resolution is resulting in substantial change in the GCAD profiles, while for the case of the EF profiles, we find very little (and almost no change in certain regions) change in the EF profiles with change in grid resolution. This indicates that the GCAD calculations are sensitive to the grid resolution of the coronal model. On the other hand, since the EF is a ratio (Equation~\ref{eqn-ef}), the effect of the change in grid resolution would affect the magnetic field term in both the numerator and the denominator, and hence the overall effect would lead to hardly any change in the ratio.

\begin{figure}
    \centering
    \includegraphics[width=0.9\linewidth]{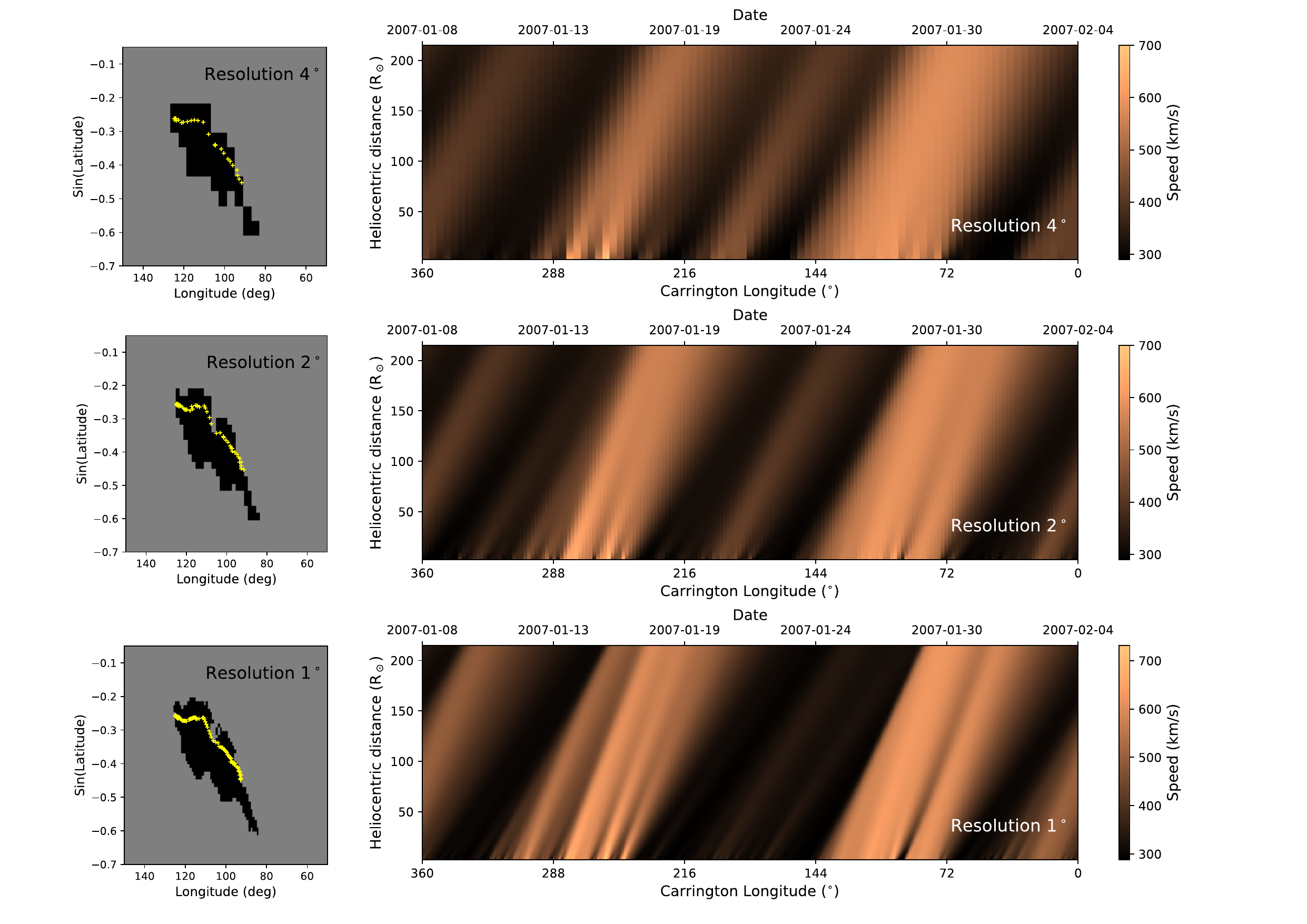}
    \caption{Top row: The modeled coronal hole (at Carrington longitude of $100^{\circ}$) for CR 2052 is shown for a grid resolution of 4$^{\circ}$ on the left, and on the right the corresponding WSA/HUX speed map from 2.5 R$_{\odot}$ to 215 R$_{\odot}$ (or 1 AU) is shown. Similarly we show the modeled coronal hole boundaries and the corresponding WSA/HUX speed maps for a grid resolution of 2$^{\circ}$ and 1$^{\circ}$ in the middle and bottom row, respectively. }
    \label{fig-ch-hss}
\end{figure}

To further understand the effect of grid resolution, in Figure~\ref{fig-ch-hss} (left) we show the modeled coronal hole boundaries of the coronal hole located at a Carrington longitude of $100^{\circ}$ for CR 2052 (see bottom panel of Figure~\ref{fig-res-ef-dmap}) for a grid resolutions of $4^{\circ}$, 2$^{\circ}$ and $1^{\circ}$, and the corresponding WSA/HUX speed maps from R$_{ss}$ (in this case 2.5 R$_{\odot}$) to 215 R$_{\odot}$ (right). The black areas on the left indicate the modeled open field region (for the above mentioned coronal hole) while the background gray region indicate the closed field regions, and the yellow crosses over-plotted inside the modeled coronal holes indicate the foot-point locations of the open field lines that are connected to the sub-Earth locations (as shown in Figure~\ref{fig-res-ef-dmap}). It can be seen on the left panel, that the foot-point locations for all the three cases, lie at the edge of the coronal hole, thus indicating that it is this boundary of the coronal hole that is producing the leading edge of the HSS (with ``leading edge'' we mean the first edge of the HSS that hits the Earth). In the right panel one can see that with change in resolution, there is a distinct shift in the longitudinal positions of the HSS. For the first HSS, the position of the HSS shifts towards the left for higher resolutions and for the second HSS, it shifts towards the right side. Thus a shift in the position of the HSS would imply a shift in the location of the leading edge of the HSS (the left most bright  edge for either of the two HSS that marks the start of the HSS for each), and hence a change in the starting time of the HSS. This is reflected as a shift in $t_{start}$ at L1 in Figure~\ref{fig-res-vr}. Additionally, we also note that with improved resolution, the widths (and hence the durations, $t_{dur}$) of the two HSSs are also changing. For the HSS on the left, the width increases with better resolution, while for the right, the width decreases with better resolution. Thus, instead of having a monotonic effect on the two widths of the two HSSs, we find that a better resolution essentially leads to better modeling of the coronal hole boundaries, which eventually leads to better modeling of the HSS. In the bottom panel, we further note that the speed differences inside an HSS (as observed in Figure~\ref{fig-res-vr}) are also captured better with better resolution, as in the HSS on the left, mutually distinct bright ridges could be observed, which can also be seen as a change in the peak speeds reached during the duration of an HSS. Thus Figure~\ref{fig-ch-hss} portrays on one hand, the close connection between coronal hole boundaries and the starting time and durations of HSSs and, on the other hand, the important role of the model resolution in connecting these two aspects effectively. 

\begin{figure}
    \centering
    \includegraphics[width=0.7\linewidth]{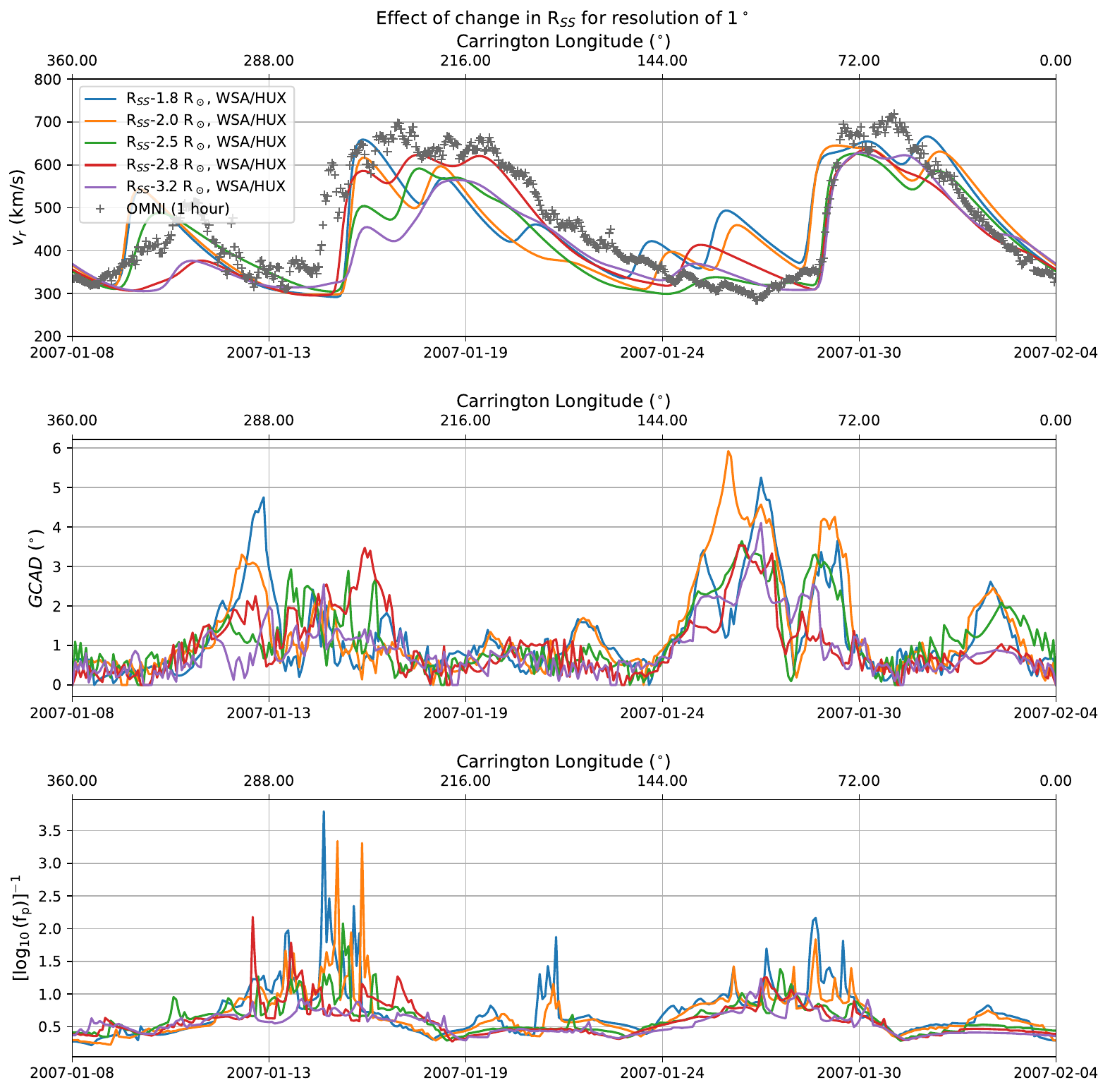}
    \caption{Effect of change in source surface height for a grid resolution of $1^{\circ}$ for CR 2052 on the solar wind speed at L1 (top panel), and the GCAD profile (middle panel) and EF profile (bottom panel) at the source surface height at the sub-Earth location.}
    \label{fig-rss-ef-dmap}
\end{figure}

\subsection{Effect of Source Surface Height}
In Figure~\ref{fig-res-vr}, we found that by using a better grid resolution, the HSS are modeled better, in particular, errors in $t_{start}$ and $t_{dur}$ could be minimized. Searching for additional improvements, we keep the grid resolution fixed at $1^{\circ}$ and investigate the effect of changing the source surface height R$_{ss}$ on our solar wind modeling of CR 2052. We change R$_{ss}$ between 1.8 - 3.2 R$_{\odot}$ and show the results in Figure~\ref{fig-rss-ef-dmap}. We first note that the solar wind speed profiles (top panel) show substantial changes for different source surface heights,  which supports the idea that the source surface height modulates the solar wind speed measured at L1. Moreover, improvement in the modeled $v_{peak}$ can also be seen. Looking into the near-Sun GCAD profiles and the EF profiles (in the middle and bottom panels, respectively), we find that both quantities are sensitive to the choice of the source surface height (unlike the case in Figure~\ref{fig-res-ef-dmap} where we found the effect of change in resolution was reflected primarily in the GCAD profile, while the EF profile hardly changed). Since both GCAD and EF are closely linked to the ambient coronal magnetic environment, any change in source surface height that alters these quantities suggests a corresponding modification in the modeled coronal magnetic environment. 

\begin{figure}
    \centering
    \includegraphics[width=0.8\linewidth]{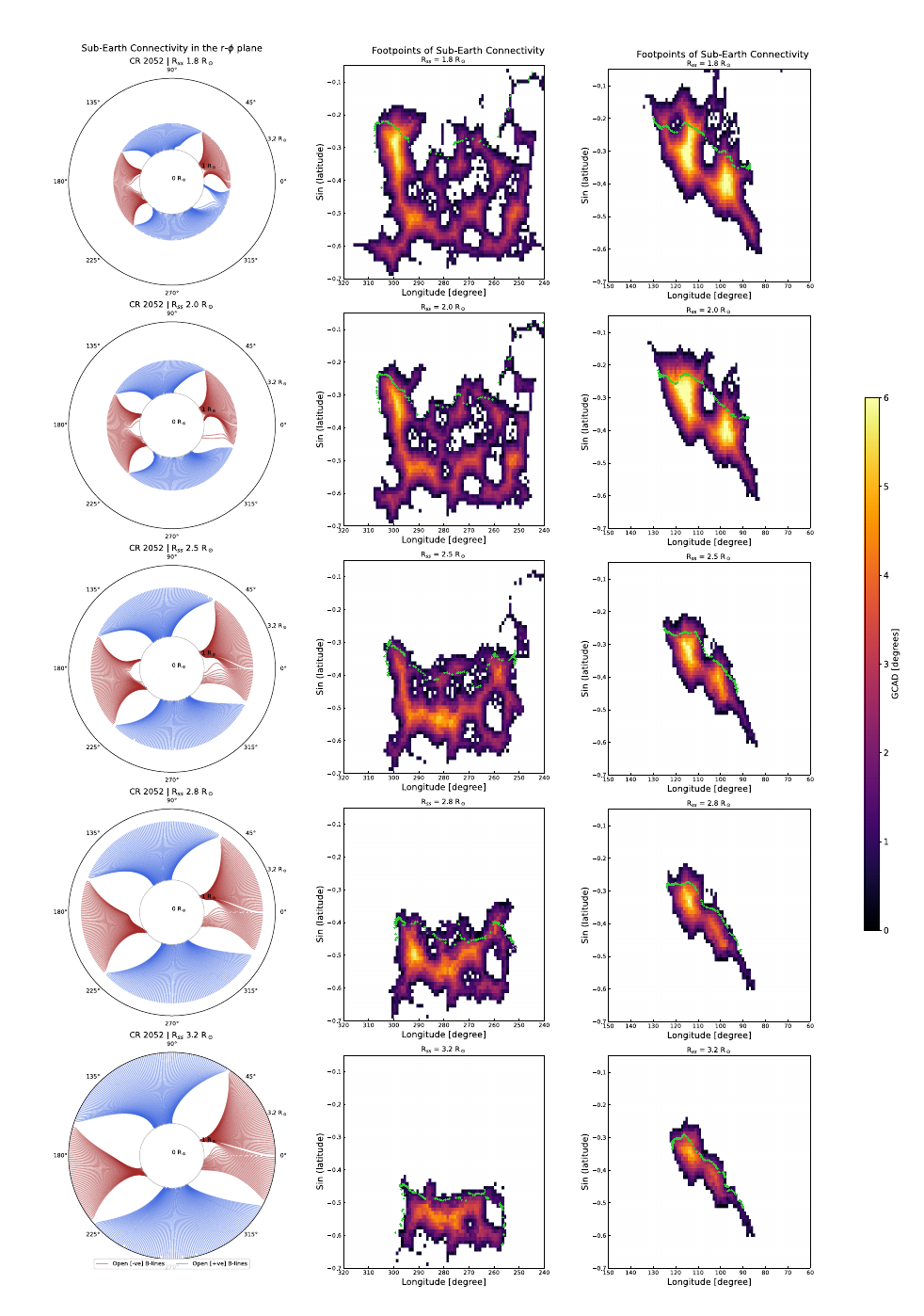}
    \caption{Left column: a polar view of the open field line connectivities for CR 2052 in the radial azimuthal ($r,\phi$) plane. The red and blue lines indicate lines of opposite polarities. The source surface height is indicated on the top of each plot, the Carrington longitudes are plotted alongside with the different radial distances marked in terms of R$_{\odot}$. Middle and right column: GCAD maps for the two coronal holes that produced the HSSs in CR 2052 for different values of the source surface height. The yellow crosses indicate the foot-point locations of the open field lines connecting the sub-Earth locations to the coronal holes.}
    \label{fig-r-phi}
\end{figure}

To gain better insight we now study in detail the coronal magnetic morphology and coronal holes that produced the HSSs for CR 2052 in Figure~\ref{fig-r-phi}. For different values of the source surface height (from $1.8R_{\odot}-3.2R_{\odot}$ as shown in Figure~\ref{fig-rss-ef-dmap}), we plot in the left column the open magnetic field lines that reach the sub-Earth locations in the radial azimuthal ($r,\phi$) plane (with red and blue depicting the 2 polarities as in Figure~\ref{fig-res-ef-dmap}), and in the middle and the right panels, we show the GCAD maps (calculated at 1R$_{\odot}$) for the two coronal holes that were responsible for producing the two HSSs in CR 2052 (bottom panel in Figure~\ref{fig-res-ef-dmap}). On each GCAD map of the coronal holes, the foot-points of the field-lines connecting to the sub-Earth locations are plotted as green crosses. On the left column, three distinct magnetic morphologies can be noted, the almost radial open field lines extending outwards, and the non-radial streamer structures which can be sub-divided into two categories, the helmet-streamers which are bi-polar (originating from coronal holes of opposite polarities) and pseudo-streamers which are uni-polar (originating  from coronal holes of the same polarity). As the source surface height is increased, the cusp heights of the pseudo-streamers remain unchanged, while the helmet streamer cusp height is always at R$_{ss}$ and therefore  keeps increasing with increasing R$_{ss}$, which is essential since helmet streamer cusps are associated with current sheets. We also notice with increasing R$_{ss}$, a pseudo-streamer (at a longitude of $180^{\circ}$) that was separating two helmet streamers is gradually giving way to two helmet streamers. On the other hand, at longitude of $0^{\circ}$, a pseudo-streamer is formed in between two helmet streamers. This clearly demonstrates that changing the source surface height drastically changes the topology of the corona like for instance the appearance or disappearance of pseudo-streamers that influence the near sun solar wind speed map (for a discussion, see \cite{Tokumaru2024SoPh}. 
To understand why an improvement in $v_{peak}$ is achieved by changing R$_{ss}$, we look into the GCAD maps of the two coronal holes in the middle and the right columns of Figure~\ref{fig-r-phi}. As expected we find that the GCAD maps show greater angular distances towards the central parts of the coronal holes, as compared to the regions at the edges of the coronal holes. We find that with change in the source surface heights, the locations of the open field-line foot-points are also changing, and as a result, the GCAD values associated with the foot-point locations on the coronal holes also change. This change in the GCAD values along the foot-points of sub-Earth connectivities, produces a change in the near Sun speed map at the source surface height via the WSA relation (Equation~\ref{eqn-v_wsa}), which finally gets propagated to L1 through the HUX model. 

To further explore the implications of this, we compare the first HSS in 
Figure~\ref{fig-rss-ef-dmap} with the GCAD map in the middle panel that studies the coronal hole that produced this HSS. For R$_{ss}$ = 1.8 R$_{\odot}$, we find that the foot-points located at the left side of the coronal hole are associated with higher GCAD values, while the right edge of the coronal hole where the foot-points are located, starts getting fragmented from a longitude of 290$^{\circ}$ onward and the foot-point locations get associated with smaller GCAD values. This gets reflected into the modeled speed profile in Figure~\ref{fig-rss-ef-dmap} as we find that the corresponding modeled HSS profile (in blue) matches the initial peak of the observed HSS profile, but it then starts descending and drifting away from the rest of the observed profile. As we increase the source surface height we find from Figure~\ref{fig-r-phi} that the fuzziness of the coronal hole boundary goes down, leading to the foot-point locations getting associated with higher GCAD values, which is in turn reflected in Figure~\ref{fig-rss-ef-dmap}, with a rise of the modeled speed profile, getting closer to the observed profile of this HSS. At a source surface of 2.8 R$_{\odot}$, the speed profile comes closest to the observed profile. When the source surface is increased further (see the speed profile for R$_{ss}\,=\,$3.2 R$_{\odot}$), no additional improvement is found, as the speed profile drops down again to lower speeds throughout the duration of the HSS. This is due to a substantial decrease in the overall size of the coronal hole (although with smoother boundary), and thus with almost similar foot-point locations as was there for R$_{ss}$ of 2.8 R$_{\odot}$, the associated GCAD values have reduced, resulting in an HSS profile with lower peak speeds. Thus, we find that one possible physical reason for the change in the solar wind speed at L1 due to a change in the source surface height, is due to the shift in the field-line foot-point location at the respective coronal hole which produce the HSS. The shifted field line foot-points being therefore responsible for sampling different speed maps near Sun and near Earth.


\begin{figure*}[ht!]
\gridline{\fig{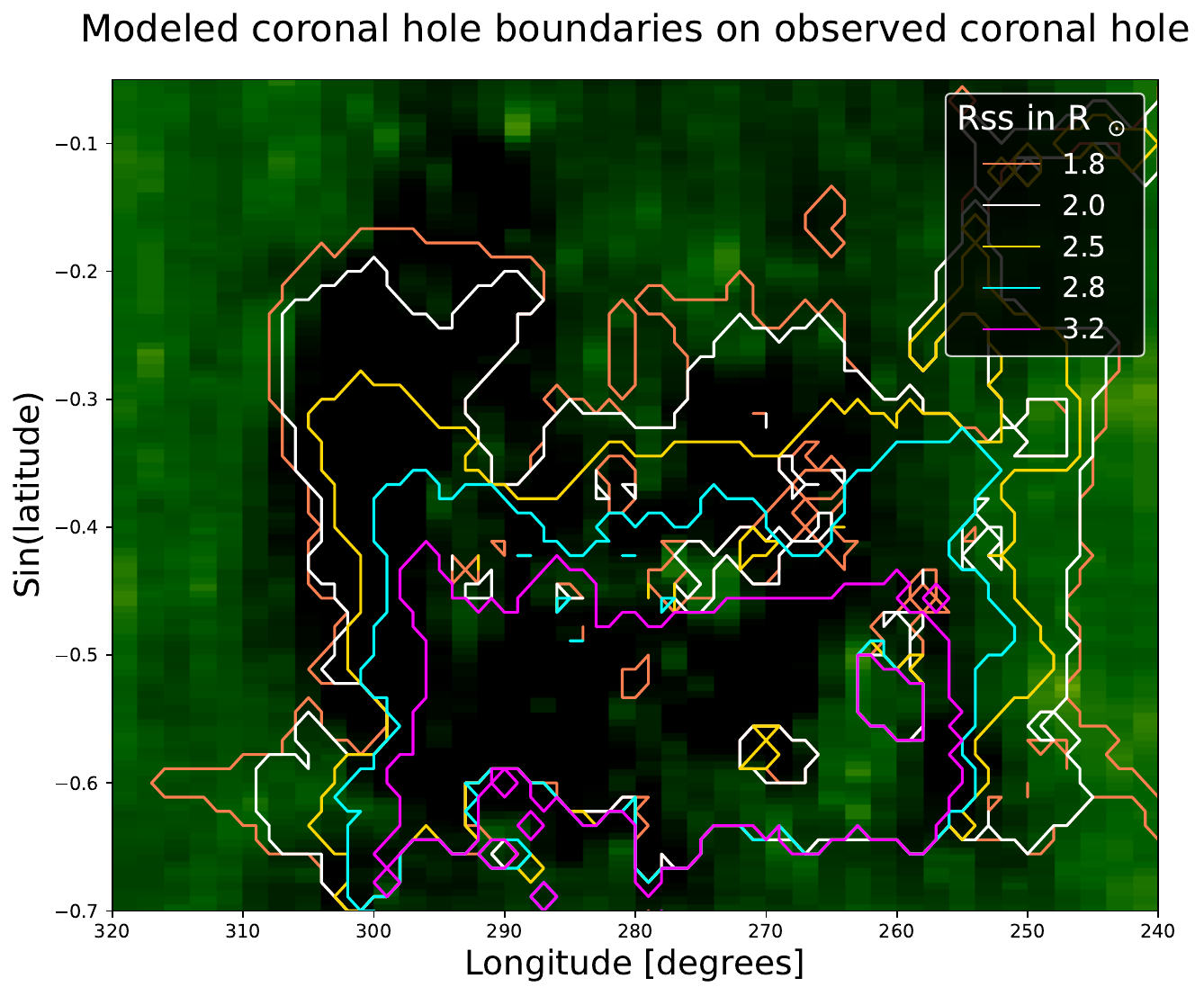}{0.4\textwidth}{(a)}
          \fig{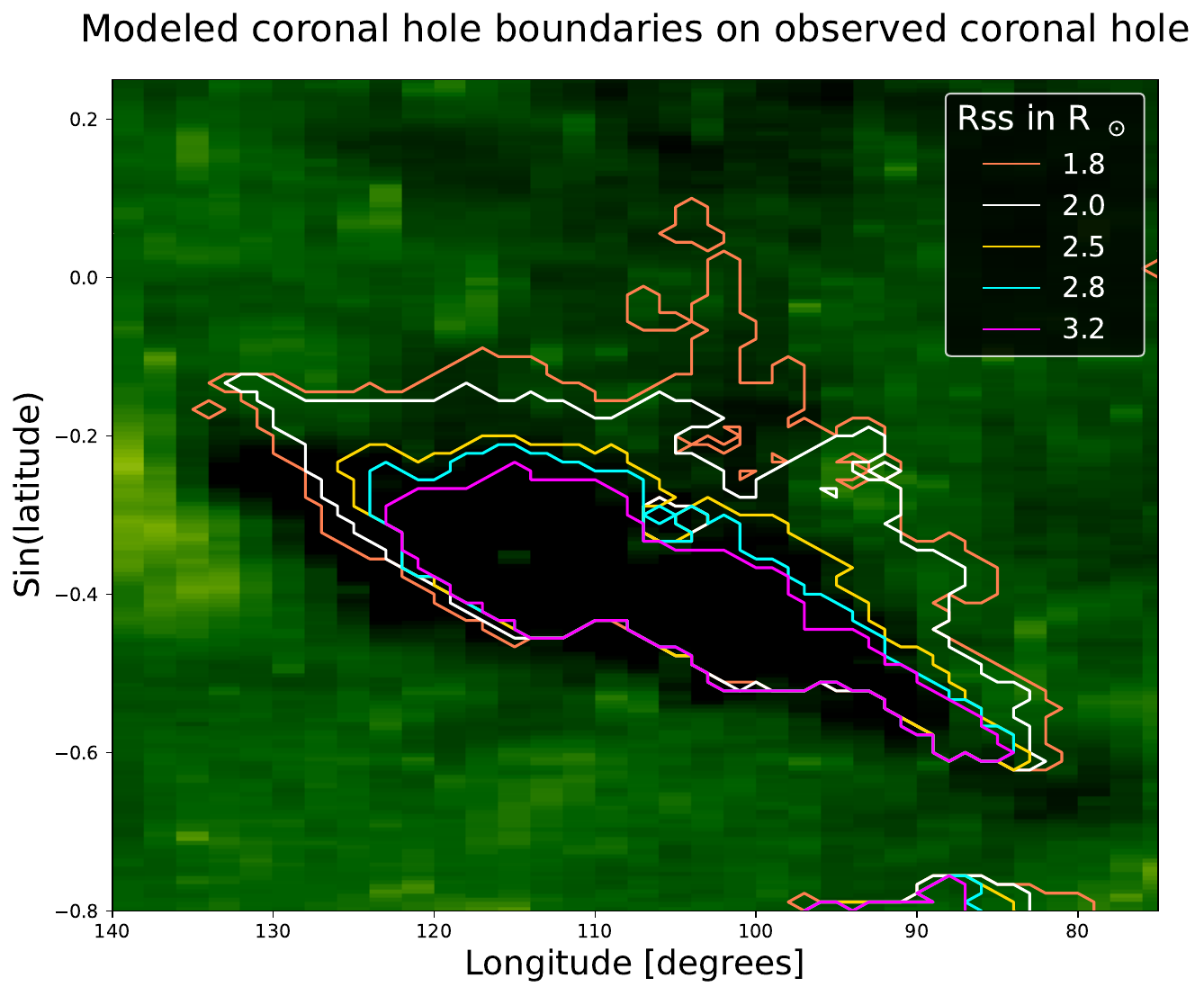}{0.4\textwidth}{(b)}
          }
\gridline{\fig{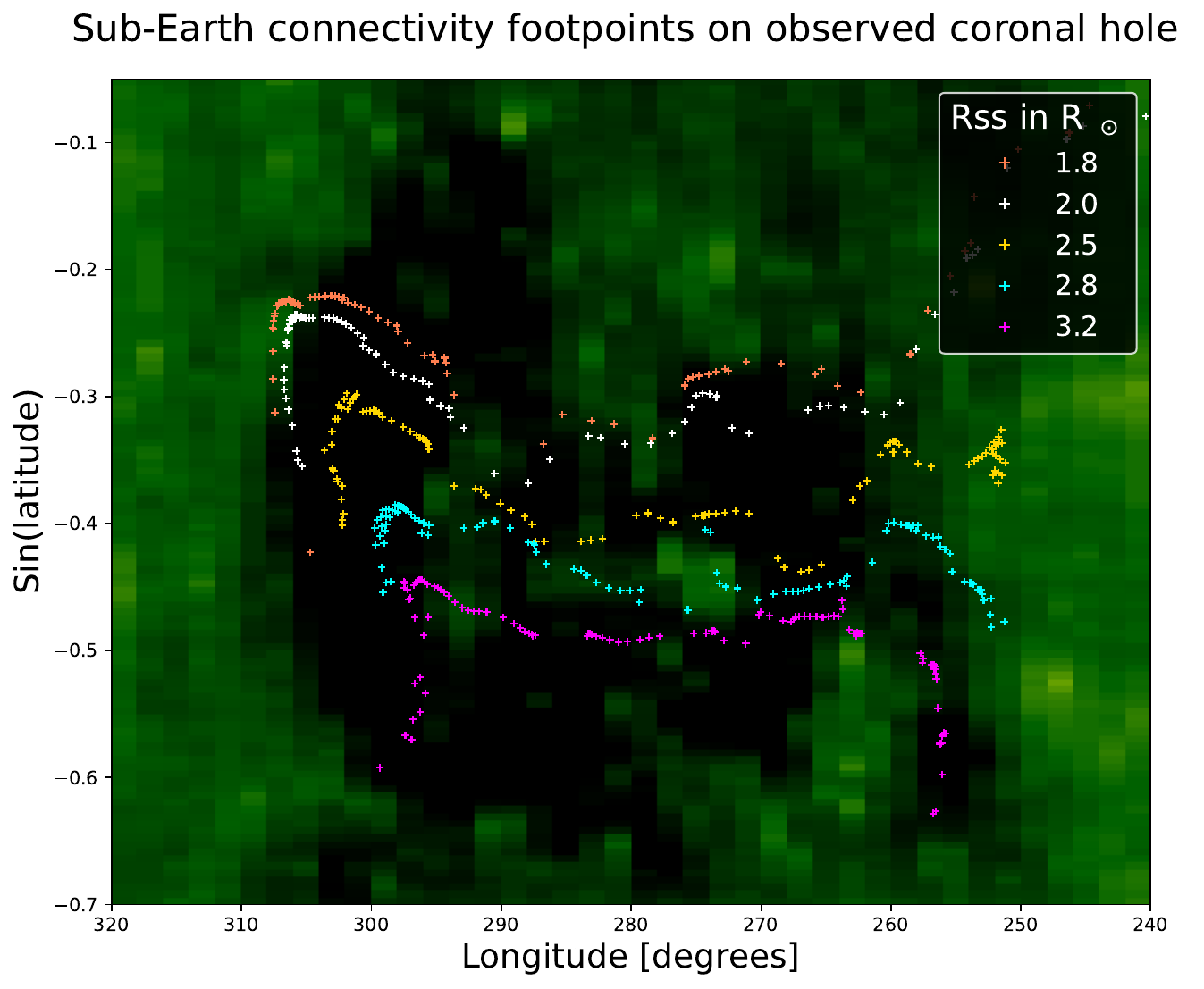}{0.4\textwidth}{(c)}
          \fig{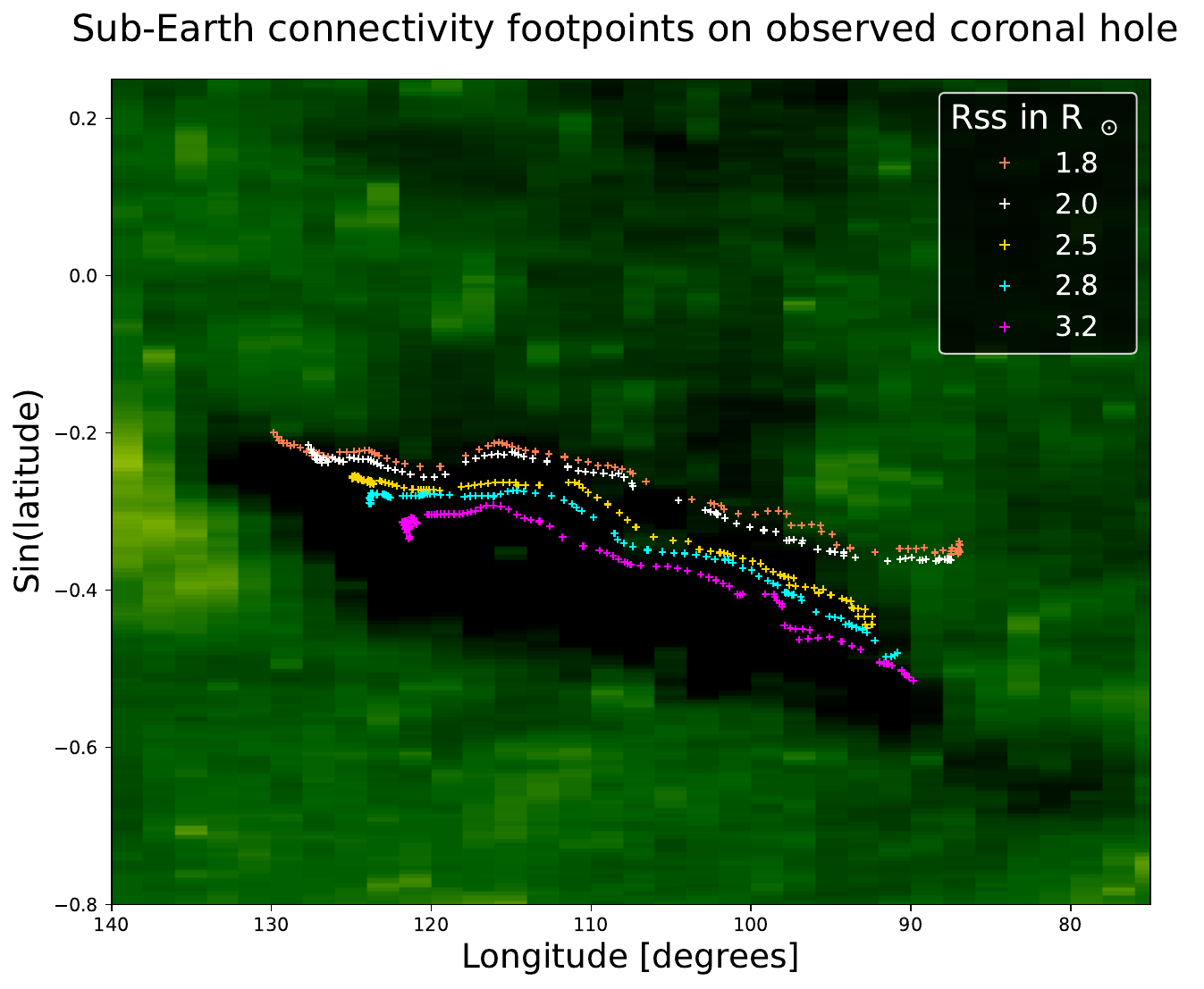}{0.4\textwidth}{(d)}
          }
\caption{Top panel a and b: coronal hole observations from SoHO/EIT 195 \AA\ for the two coronal holes in CR 2052. The modeled coronal hole boundaries for different source surface heights are overlaid. Bottom panel c and d: observations of the same coronal holes with the foot-point locations of the open field lines that connect the sub-Earth locations to the coronal holes. Different colors indicate foot-points associated with different R$_{ss}$.}\label{fig-ch-cont}
\end{figure*}

However, we emphasize that this change in the foot-point positions is also governed by a change in the modeled coronal hole boundaries (which is leading to a change in the size of the modeled CH) when imposing different source surface heights. Therefore, it becomes important to constrain the size (or boundaries) of the modeled coronal holes, which in turn would limit the choice of R$_{ss}$. We employ coronal hole observations to validate our modeled coronal holes (obtained by varying R$_{ss}$). In Figure~\ref{fig-ch-cont}(a) and (b) we overlay the modeled coronal hole boundaries for the different source surface heights on top of the observations from SOHO/EIT 195 \AA\ for the two coronal holes that were responsible for producing the two HSSs at L1 in CR 2052. In Figure~\ref{fig-ch-cont}(c) and (d) we overlay the foot-point locations of the open field lines (for the different R$_{ss}$) that are connected to the sub-Earth locations on the observed coronal holes. We can see in the top panel that the larger R$_{ss}$, the smaller the derived CH, and the shift in the foot-point locations can be seen in the bottom panel. We find that by changing R$_{ss}$ the model can essentially capture the different parts of the coronal hole: for example, in the left panel, only source surface heights lower than 2.5 R$_{\odot}$ are able to capture the upper left portions of the observed coronal hole that are left out by the contours obtained from higher source surface heights, which leads to different modeled $v_{peak}$ for the first HSS as discussed earlier. We have already seen that including this part of the coronal hole was essential in order to accurately model $v_{peak}$ of the resulting HSS, the inclusion of higher values of R$_{ss}$ is important to accurately model the peak speeds reached throughout the duration of the HSS. It should be noted that we do not intend to select any particular modeled coronal hole boundary that seems to be a best match with the observed coronal holes. Instead, we intend to demonstrate the usefulness of coronal hole observations as an observational constraint on the modeled solar wind profiles at L1.  In other words, in order to match the solar wind observations at L1, we have changed R$_{ss}$, which in turn leads to different sizes of the modeled coronal holes. We now use coronal hole observations to ensure that the modeled coronal holes corresponding to the different solar wind solutions, don’t overestimate the observed coronal hole sizes, as it often happens for low R$_{ss}$ values \citep[see e.g.~][]{Linker2017ApJ}. This in turn will help to ensure that we do not use any unphysical value of R$_{ss}$ in order to match the solar wind condition at L1. Thus, the observed coronal hole sizes can be used as a lower limit to truncate R$_{ss}$.
 
We repeat the above experiment for different CRs (that correspond to different phases of solar cycle 23 and 24), with different values of R$_{ss}$ (in the range of 1.6-3.3 R$_{\odot}$) and the corresponding root mean squared errors (RMSEs) for each case are provided in Table~\ref{tab1}. It can be seen that the values of R$_{ss}$ that best matches observations (i.e.~with minimum RMSEs), are different for different CRs and for the studied events, they range between 1.8-3.1 R$_{\odot}$. We do not notice any particular solar cycle trend in the adjusted R$_{ss}$ values that best matches with observations, however, we find different values for different CRs. So, our results indicate that firstly, it is not always possible for a single R$_{ss}$ value to capture all the different important aspects of an HSS, and secondly, it is also not straightforward to predict a single R$_{ss}$ value for any particular CR. Thus, in future studies, a possible strategy could be considering an ensemble of solar wind solutions at L1 for different values of R$_{ss}$ (similar to the top panel of Figure~\ref{fig-rss-ef-dmap}), together with an ensemble of corresponding modeled coronal hole boundaries that are constrained by coronal hole observations (similar to Figure~\ref{fig-ch-cont}).
This will on the one hand provide a physical range of possible solar wind profiles (for different values of R$_{ss}$) and, on the other hand, it would ensure that the ensemble of solar wind speed profiles are constrained by available coronal hole observations. This way one can make sure that only those solar wind solutions are considered in the ensemble for which the corresponding coronal hole boundaries do not exceed the observed boundaries.

\begin{longrotatetable}
\begin{deluxetable*}{ccccccccccccccccccc}  
\centering
\setlength{\tabcolsep}{1.2pt}
\tablecaption{The root mean squared errors (RMSE) for the different CRs studied in this work are tabulated below for the various source surface heights (R$_{ss}$) ranging between [1.6 - 3.3 R$_{\odot}$]. \label{tab1}}
\tablecolumns{19}
\tablenum{1}
\tablewidth{0pt}
\tablehead{
\colhead{Carrington} & 
\colhead{RMSE} & 
\colhead{RMSE} & \colhead{RMSE} & \colhead{RMSE} & \colhead{RMSE} & \colhead{RMSE} & \colhead{RMSE} & \colhead{RMSE} & \colhead{RMSE} & \colhead{RMSE} & \colhead{RMSE} & \colhead{RMSE} & \colhead{RMSE} & \colhead{RMSE} & \colhead{RMSE} & \colhead{RMSE} & \colhead{RMSE} & \colhead{RMSE}\\
\colhead{Rotation} & \colhead{\footnotesize(1.6 R$_{\odot}$)} &
\colhead{\footnotesize(1.7 R$_{\odot}$)} & \colhead{\footnotesize(1.8 R$_{\odot}$)} & \colhead{\footnotesize(1.9 R$_{\odot}$)} & \colhead{\footnotesize(2.0 R$_{\odot}$)} & \colhead{\footnotesize(2.1 R$_{\odot}$)} & \colhead{\footnotesize(2.2 R$_{\odot}$)} & 
\colhead{\footnotesize(2.3 R$_{\odot}$)} & \colhead{\footnotesize(2.4 R$_{\odot}$)} & \colhead{\footnotesize(2.5 R$_{\odot}$)} & \colhead{\footnotesize(2.6 R$_{\odot}$)} & \colhead{\footnotesize(2.7 R$_{\odot}$)} & \colhead{\footnotesize(2.8 R$_{\odot}$)} & 
\colhead{\footnotesize(2.9 R$_{\odot}$)} & \colhead{\footnotesize(3.0 R$_{\odot}$)} & \colhead{\footnotesize(3.1 R$_{\odot}$)} & \colhead{\footnotesize(3.2 R$_{\odot}$)} & \colhead{\footnotesize(3.3 R$_{\odot}$)}
}
\startdata
2052 & 96.64 & 89.52 & 93.38 & 90.99 & 95.52 & 98.44 & 97.25 & 85.43 & 73.02 & 77.56 & 63.28 & 71.78 & 66.12 & 87.01 & 83.99 & 82.77 & 87.57 & 96.07 \\ 
    2060 & 110.05 & 101.46 & 96.81 & 95.82 & 91.17 & 85.38 & 82.64 & 68.76 & 89.25 & 88.33 & 83.33 & 68.52 & 81.14 & 83.55 & 101.44 & 105.09 & 121.92 & 117.89 \\ 
    2081 & 114.99 & 109.74 & 77.54 & 101.42 & 95.97 & 92.99 & 82.73 & 71.84 & 88.06 & 58.97 & 68.95 & 84.24 & 94.29 & 84.15 & 86.84 & 89.72 & 74.35 & 94.03 \\ 
    2093 & 102.94 & 99.71 & 76.96 & 61.29 & 59.7 & 65.22 & 53.36 & 53.59 & 48.1 & 43.8 & 40.02 & 55.18 & 54.5 & 56.48 & 60.74 & 63.87 & 60.73 & 62.68 \\ 
    2098 & 151.56 & 134.21 & 128.86 & 118.44 & 100.35 & 81.16 & 75.11 & 69.09 & 59.56 & 72.02 & 78.18 & 79.05 & 64.1 & 60.21 & 55.95 & 43.85 & 48.03 & 49.75 \\ 
    2100 & 117.89 & 115.67 & 113.77 & 107.86 & 101.1 & 95.13 & 92.84 & 98.16 & 96.68 & 95.66 & 84.63 & 87.9 & 96.51 & 95.07 & 88.08 & 89.72 & 89.31 & 85.29 \\ 
    2104 & 131.11 & 127.24 & 120.27 & 118.48 & 124.82 & 123.75 & 116.85 & 111.65 & 105.29 & 95.66 & 82.54 & 61.86 & 56.15 & 57.53 & 57.08 & 62.24 & 63.72 & 54.72 \\ 
    2108 & 103.9 & 103.55 & 105.84 & 99.78 & 101 & 101.63 & 107.62 & 104.15 & 100.4 & 101.65 & 97.05 & 94.95 & 96.23 & 98.06 & 100.55 & 103.17 & 100.18 & 104.98 \\ 
    2114 & 81.31 & 87.12 & 78.62 & 75.58 & 75.77 & 77.67 & 84.34 & 87.17 & 95.4 & 96.4 & 100.69 & 104.23 & 101.86 & 104.58 & 102.71 & 103.78 & 101.03 & 102.06 \\ 
    2123 & 78.7 & 88.6 & 84.14 & 86.12 & 87.67 & 78.77 & 85.76 & 88.23 & 83.66 & 92.1 & 95.38 & 92.15 & 93.63 & 98.34 & 98.68 & 92.95 & 98.98 & 95.84 \\ 
    2124 & 91.17 & 83.27 & 81.69 & 82.9 & 83.46 & 85.02 & 79.54 & 83.68 & 82.67 & 91.69 & 91.79 & 94.65 & 92.88 & 92 & 91.25 & 87.82 & 87.67 & 91.26 \\ 
    2138 & 131.95 & 120.59 & 104.78 & 88.98 & 88.74 & 79.63 & 68.05 & 59.91 & 55.14 & 56.74 & 54.99 & 58.08 & 65.56 & 60.64 & 62.35 & 61.75 & 73.2 & 76.19 \\ 
    2185 & 157.8 & 161.34 & 155.3 & 139.79 & 140.61 & 150.51 & 147.04 & 146.39 & 144.82 & 141.12 & 146.19 & 146.24 & 144.73 & 145.95 & 134.91 & 140.64 & 138.93 & 139.46 \\ 
    2202 & 129.72 & 115.02 & 105.17 & 111.03 & 111.21 & 103.78 & 96.08 & 81.41 & 101.33 & 70.17 & 72.08 & 78.92 & 76.53 & 56.93 & 72.3 & 71.98 & 68.29 & 78.77 \\ 
    2209 & 141.47 & 140.51 & 139.45 & 139.26 & 147.8 & 142.09 & 126.75 & 118.95 & 108.03 & 105.3 & 108.85 & 117.04 & 114.43 & 119.55 & 120.86 & 124.8 & 114.87 & 120.29 \\ 
\enddata
\end{deluxetable*}
\end{longrotatetable}

\subsection{On the relevance of CH observation for ADAPT Maps} \label{sec-adapt}

\begin{figure}
    \centering
    \includegraphics[width=0.6\linewidth]{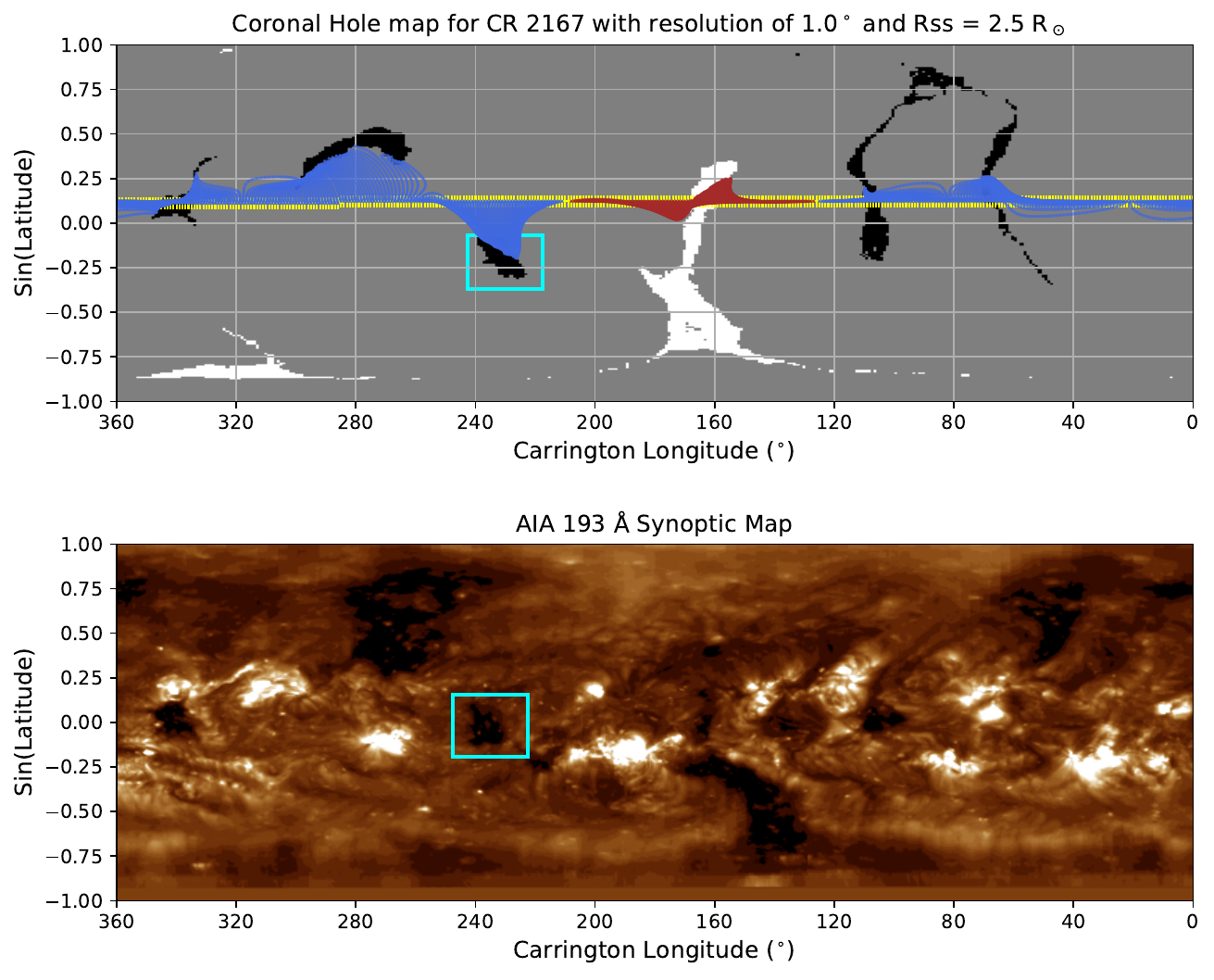}
    \caption{Top panel: the modeled coronal hole map for CR 2167. The blue and red lines show the open field lines connecting the open field regions to the sub-Earth locations (indicted by yellow crosses), where red/blue corresponds positive/negative magnetic field polarity. Bottom panel: AIA 193 \AA\ synoptic map for the same CR. The aqua rectangle plotted in the two panels indicate a particular coronal hole that is discussed in Section~\ref{sec-adapt}.}
    \label{fig-cr2167-gong}
\end{figure}

\begin{figure*}[ht!]
\gridline{\fig{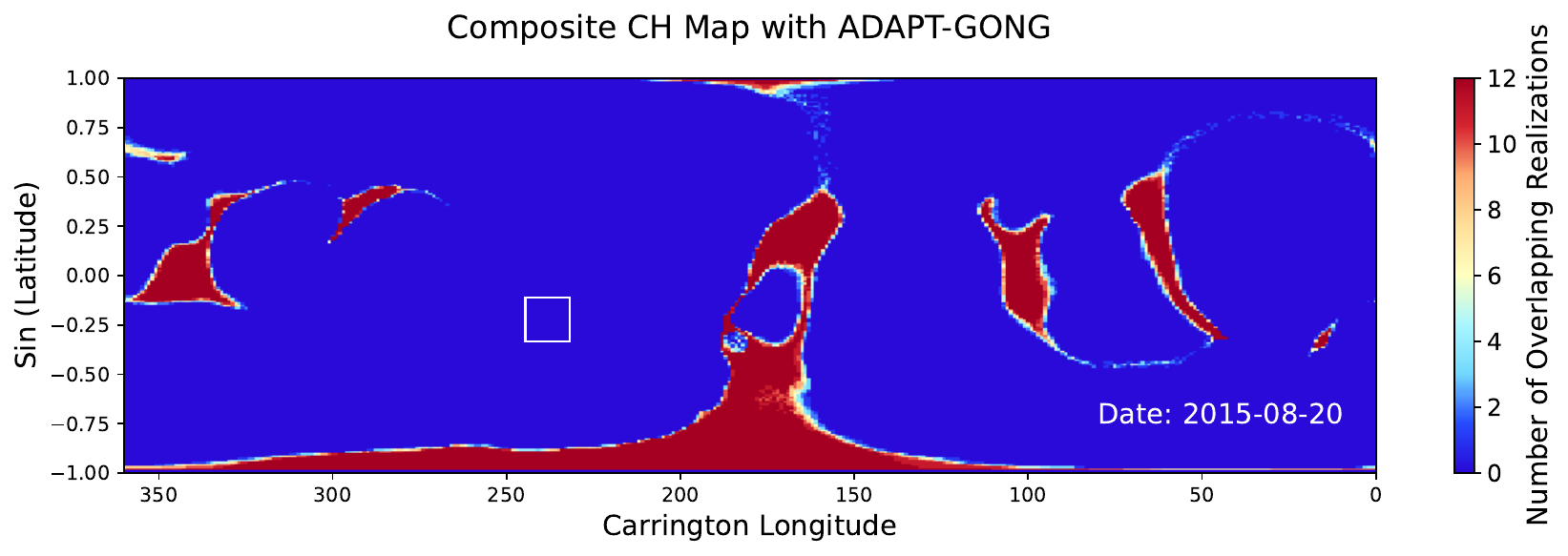}{0.8\textwidth}{(a)}
          }
\gridline{\fig{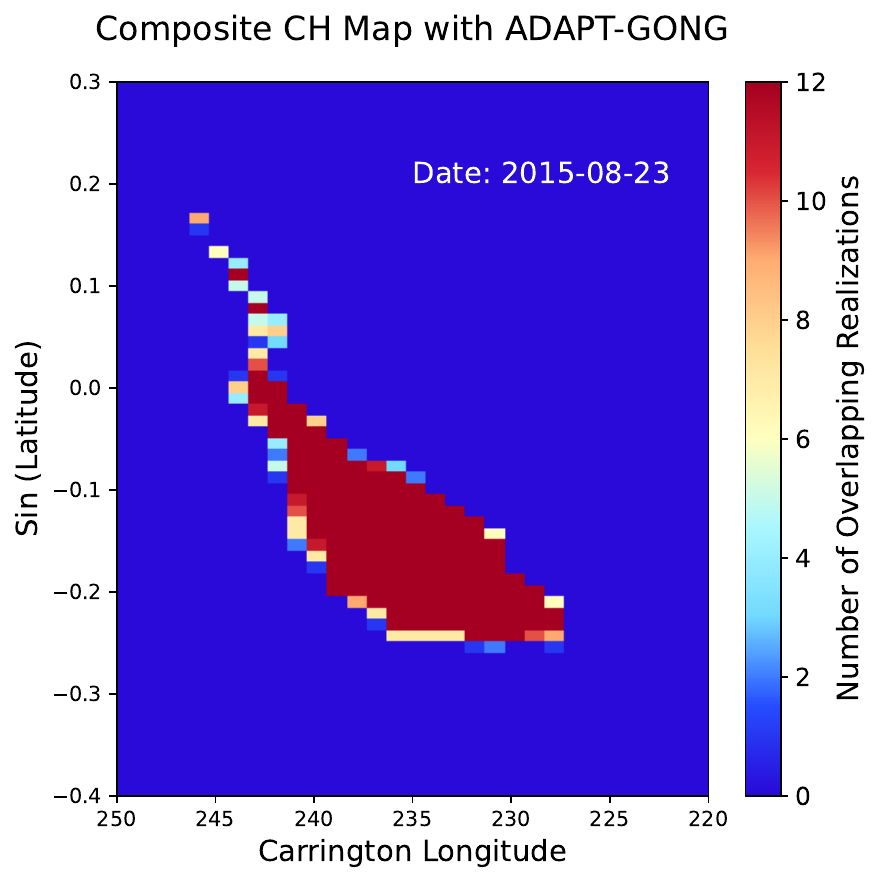}{0.3\textwidth}{(b)}
          \fig{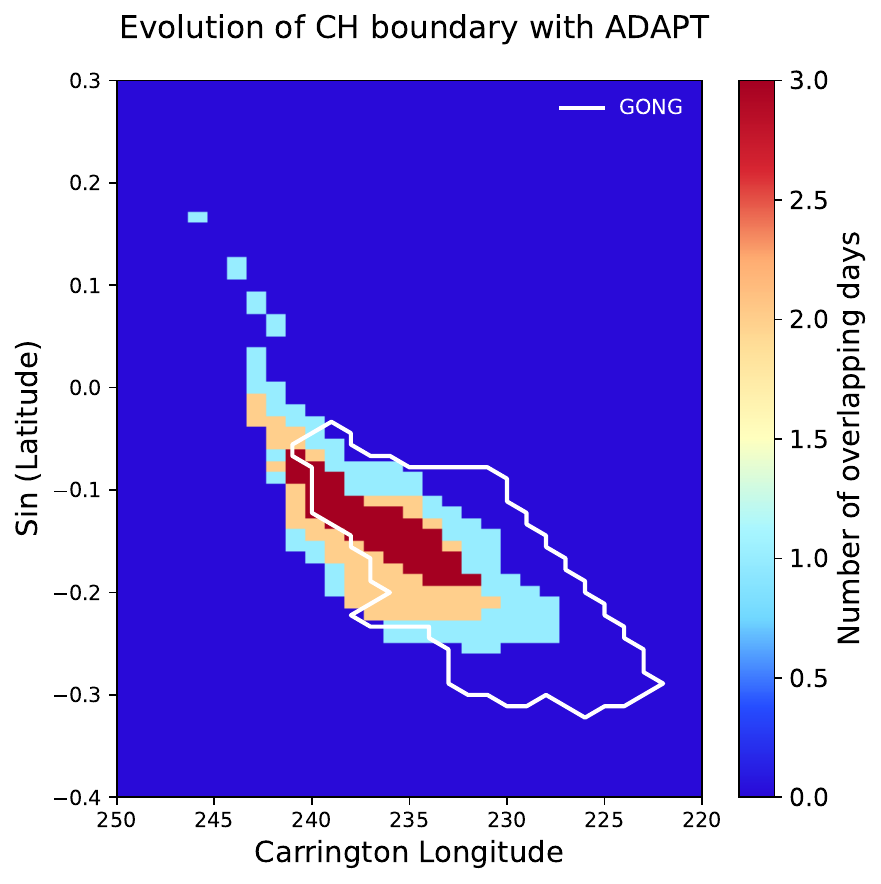}{0.3\textwidth}{(c)}
          \fig{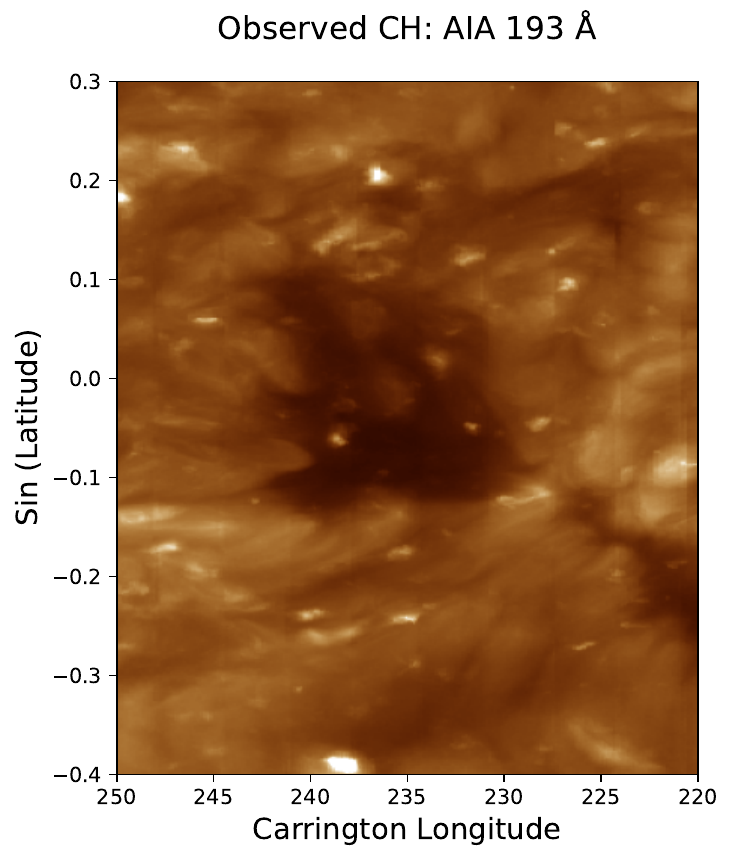}{0.255\textwidth}{(d)}
          }
\caption{Top Panel: (a) overlap image of the modeled open field regions of the 12 ADAPT ensemble maps created on 20 August 2015. Bottom panel: (b) cropped overlap image of the modeled open field regions for the 12 ADAPT ensembles created on 23 August 2015, showing the coronal hole indicated in the box in Figure~\ref{fig-cr2167-gong}, (c) cropped overlap image of the fifth realization of the ADAPT maps for three consecutive days between 21-23 August 2015. The white line contour over-plotted shows the boundary of the same coronal hole obtained by modeling with the GONG CR map in Figure~\ref{fig-cr2167-gong}, and (d) for comparison  the cropped image of the observed coronal hole in SDO/AIA 193 \AA.}\label{fig-adapt-1}
\end{figure*}

A key limitation of using traditional synoptic (CR) magnetic maps in solar wind modeling is that they are constructed over an entire solar rotation, preventing them from capturing the real-time evolution of the magnetic field. As a result, they are not suitable for real-time modeling. In contrast, quasi-synchronic maps like ADAPT make use of data assimilation techniques to provide a more time-accurate representation of the Sun’s magnetic field by blending observations with a flux transport model. However, since the ADAPT maps are produced every 2 hours (updated with new observations), a key question arises as to how to choose an appropriate date and time of an ADAPT map for a particular coronal hole under study. To investigate this question we show a modeled coronal hole map for CR 2167 using again a GONG CR map as input in the top panel of Figure~\ref{fig-cr2167-gong}. The sub-Earth connectivities are indicated by red and blue lines (different colors indicating opposite polarities). The coronal hole enclosed inside the aqua rectangle in the figure (located at a longitude of $230^{\circ}$ and Sin(Latitude) of -0.23) appeared in AIA 193 \AA\ field of view at the east limb on 15 August 2015 and was observed at the disk center on 20 August, 2015. This coronal hole produced an HSS that reached L1  \citep[discussed in detail in][]{Reiss2024ApJS}. It can be seen that the modeled coronal hole is indeed connected to the sub-Earth location. A synoptic map constructed using observations from SDO/AIA 193 \AA\ for the same CR is plotted in the bottom panel, where the corresponding observed coronal hole can be identified at a similar location (enclosed inside the aqua rectangle). Considering this coronal hole, we next run our model with the last ADAPT map created on the same day the coronal hole was observed at disk center (20 August 2015), constructed from GONG observations from the GONG-ADAPT database. Using the ADAPT maps, we create an overlapping coronal hole map for all 12 ensemble members and the resulting overlapped image is shown in Figure~\ref{fig-adapt-1}(a). It can be seen that the above mentioned coronal hole is not present in any of the 12 realizations. This missing coronal hole is particularly important because it is connected to the sub-Earth points and was producing an HSS. Since ADAPT uses observations to continuously update the magnetograms, it is essential to consider maps that were created at a later time, so as to make sure that any new flux emergence that emerged on far side is accounted for (i.e., allowed to rotate onto the near side of the Sun and taken in account). Thus, we again used ADAPT maps that were created three days later, on 23 August 2015, and again the resulting overlapped coronal hole map cropped to the location of the coronal hole of interest, is shown in Figure~\ref{fig-adapt-1}(b). For this later date the coronal hole that was missing in  Figure~\ref{fig-adapt-1}(a) is now present in all 12 realization in Figure~\ref{fig-adapt-1}(b), with very small differences in the different realizations. This raises the question on how to select the correct time for the ADAPT maps for solar wind modeling, not only for planning future science studies, but also for real-time forecasting. To shed more light on this, we again selected the last ADAPT maps (created at 20 UT) for each day starting from August 21 (day 1 after the coronal hole was observed at the disk center in AIA field of view), to August 23 (day 3), and for these set of maps, we run the model again to analyze further. The modeled coronal hole boundaries for these three days are over-plotted in Figure~\ref{fig-adapt-1}(c). Note that since we have already shown that the coronal hole is present in each of the realizations for the map created on day 3 (Figure~\ref{fig-adapt-1}(b)), for Figure~\ref{fig-adapt-1}(c), we randomly selected the fifth realization for all three days. It can be seen that firstly, the mentioned coronal hole is present in all three days, and secondly, the size of the coronal hole increases from day 1 to day 2 to day 3. For comparison, we also plot the coronal hole boundary obtained using the GONG CR map (from Figure~\ref{fig-cr2167-gong}) on the same plot (shown in the white line). We find that the modeled boundary from ADAPT maps comes closest to the boundary obtained from the GONG CR map for day 3, which is three days after the coronal hole was observed at disk center. To put the observational perspective in this comparison, we also show the same coronal hole as observed in AIA 193~\AA\ on August 20 in Figure~\ref{fig-adapt-1}(d). It should be noted that the resolution of the image in (d) is much higher than the resolution of the contours plotted in panels (b) and (c). It is also worth noting that compared to GONG, the modeled coronal hole boundaries from ADAPT (panel c) align more closely with the observed boundaries (panel d).   

Thus, in using ADAPT magnetic maps, this should be kept in mind that due to continuous updating of the flux-transport models with fresh observations, newly emerging coronal hole can take several days to get incorporated in the maps. Although it should be noted that the discussed CH was visible in AIA 193~\AA\ field of view from 15 August onwards, when it appeared at the East limb. Thus, ample care should be taken in choosing the correct ADAPT map for a particular study. As a starting point in this aspect, our results indicate that coronal hole observations can play the role of an important constraint in choosing ADAPT maps, but more detailed studies in the future on how to choose an ADAPT map for a particular study would be very useful.

\section{Discussion and Conclusion} \label{conclusion}

We study the causes of errors in modeling the ambient solar wind at L1, particularly the errors encountered during the occurrence of an HSS. With a modeling framework consisting of PFSS/WSA/HUX model, we studied 15 CRs between 2007 and 2018 and we presented a detailed analysis for CR 2052. We consider three key modeling aspects of an HSS, namely the starting time of an HSS ($t_{start}$), the duration of the HSS ($t_{dur}$) and the peak speed reached during an HSS ($v_{peak}$). We found that these three aspects are sensitive to changes in the coronal magnetic model. Thus, we conducted a detailed investigation of the WSA model settings to explain uncertainties in the aforementioned three aspects of an HSS. In the following, we discuss our main findings from this study.  

\begin{enumerate}
    \item We found that increasing the resolution of the coronal model improves the identification of coronal hole boundaries, which leads to improvements in $t_{start}$ and $t_{dur}$ of the modeled HSS profile at L1 (Figure~\ref{eqn-v_wsa}). This improvement is not monotonous, as for one HSS a higher resolution lead to an earlier $t_{start}$ and for another it lead to a later $t_{start}$ that eventually minimized the error in the modeled and observed $t_{start}$ for both HSSs. We repeated the exercise for other CRs (that included one or two HSSs) and found similar improvements in the modeled $t_{start}$ and $t_{dur}$. To understand this improvement better
    we show in Figure~\ref{fig-res-ef-dmap} that by changing the resolution, the near-Sun GCAD profile at the Sub-Earth locations change significantly (due to changes in resolution that alters the open-closed boundaries, and hence the GCAD distances). On the contrary, we found that the near-Sun expansion factor profile at the sub-Earth locations remains mostly unchanged with change in the grid resolution. This makes sense because the expansion factor is a ratio, and hence the change in resolution would affect both the numerator and the denominator. In Figure~\ref{fig-ch-hss}, we further demonstrate how a better grid resolution affects the modeled HSS map from R$_{ss}$ to 1 AU, leading to improvements in key aspects like $t_{start}$ and $t_{dur}$ of the HSS. We find our results in agreement with \cite{Riley2013JGRA}, who also reported on improvement in the correlation coefficient with higher model resolution using the CORHEL model suite.

    \item In Figure~\ref{fig-rss-ef-dmap}, we showed that optimizing the source surface height further improves the modeled solar wind speed at L1. Previous studies have already demonstrated that varying the source surface height can lead to better agreement of model output with observations for different physical parameters like the interplanetary magnetic field, coronal hole sizes, the open magnetic flux \citep[see; ][]{Lee2011SoPh,Arden2014JGRA,Nikolic2019SpWea,Asvestari2019JGRA}. In this study, we investigated why a changing R$_{ss}$ leads to a substantial change in the solar wind profile at L1. In Figure~\ref{fig-rss-ef-dmap} we can see that a change in R$_{ss}$, changes both the near-Sun GCAD and EF profiles, thus reflecting a change in the modeled coronal magnetic topology. This was supported by the left column of Figure~\ref{fig-r-phi}, where we show that for different $R_{ss}$, the magnetic structures undergo significant changes between pseudo-streamers and helmet streamers. As has been pointed out in earlier studies helmet streamers and pseudo-streamers can influence the solar wind speed profiles \citep[see ][]{Wang2007ApJ,Riley2012SoPh,Wallace2020ApJ,Tokumaru2024SoPh}. Thus in future work, it would be valuable to assess the effect of these changing magnetic structures from helmet streamers to pseudo-streamers on the modeled solar wind profile.

    \item Exploring the effect of the R$_{ss}$ further, we found that by changing R$_{ss}$, the boundaries of the modeled coronal holes undergo significant changes which in turn, changes the sizes of the modeled coronal holes \citep[as reported earlier by][]{Asvestari2019JGRA,Nikolic2019SpWea} which can have important implications on the inferred physics. \cite{Linker2017ApJ} analyzed modeled coronal hole sizes for different R$_{ss}$ values and reported on a clear discrepancy between the open flux from the modeled coronal hole areas and the open flux from in-situ measurement. \cite{Riley2019ApJ} also studied modeled coronal hole sizes for different R$_{ss}$ and reported that this `open flux problem' may be resolved (at least in part) by the addition of a modest polar flux to photospheric magnetic field maps. In the context of solar wind modeling, in this study, we find that besides changing the modeled coronal hole sizes, a change of the coronal hole boundaries also changes the GCAD values associated with the different parts of the coronal hole (middle and right columns of Figure~\ref{fig-r-phi}). Further, different values of R$_{ss}$ also change the location of the open-field-line footpoints at the coronal holes. It is then the different GCAD values that are associated with the foot-point locations of the field lines connected to sub-Earth locations, that get sampled through the WSA model to produce different speed profiles at L1 for different R$_{ss}$. We also showed that the differing foot-point locations for differing R$_{ss}$, explains the modeling of important aspects of an HSS like the multiple peak speeds attained during an HSS. \cite{Meadors2020SpWea} reported earlier on the improvement in solar wind prediction that is achieved by adjusting R$_{ss}$ using a data assimilation with particle filtering approach. This work provides a detailed explanation of why a change in R$_{ss}$ leads to improvement in the predicted solar wind profile. It is interesting to note that \cite{Issan2023SpWea} used a Bayesian inference and global sensitivity analysis and found the R$_{ss}$ to play a less critical role as compared to other parameters in the WSA model. In this study, we did not vary any of the WSA parameters that appear in Equation~\ref{eqn-v_wsa} like in our previous study in \cite{Reiss2020ApJ}, to focus on the effect of varying the model grid resolution and the source surface height. To further ensure that our modeled coronal holes, and hence the choice of R$_{ss}$ are physically constrained, we validate the modeled boundaries of the coronal holes (for the different source surface heights) with coronal hole observations from SOHO/EIT 195~\AA\ (Figure~\ref{fig-ch-cont}(a) and (b)). By overlaying the footpoint locations of the sub-Earth connectivities on the observed coronal holes (Figure~\ref{fig-ch-cont}(c) and (d)), we also showed how different R$_{ss}$ essentially leads to capturing different regions of a coronal hole. In this context, it should also be noted that there are various factors that lead to uncertainties in the observed coronal hole boundaries in EUV images \citep{Reiss2024ApJS} which are not taken into account in this study. 

    \item For CR 2052, we found the best agreement with the observed speed at L1 for R$_{ss}$ of 2.8 R$_{\odot}$ (see Table~\ref{tab1}). We repeated the exercise for other CRs with HSSs, and found different values of $R_{ss}$ (ranging between 1.8 - 3.1 $R_{\odot}$) that led to the best agreement with the observed speed at L1 (Table~\ref{tab1}). These values of R$_{ss}$ overlap considerably with the results from previous studies \citep{Lee2011SoPh,Arden2014JGRA,Nikolic2019SpWea,Asvestari2019JGRA,Meadors2020SpWea}, although it should be noted that R$_{ss}$ was varied in some of these studies with different aims (like modeling the open field regions, or the open magnetic flux). However, our results also showed that it is not possible for a single R$_{ss}$ to capture all the essential aspects of an HSS. As R$_{ss}$ is a free parameter that can be tuned, extending our methodology to a large dataset of CRs covering the entire solar cycle would be valuable to understand if there is any clear relationship between the phase of the solar cycle and the value R$_{ss}$ that best agrees with solar wind observation. However, through this current study, we also indicate that this free parameter should be varied in a physically constrained paradigm, and that coronal hole observations should be considered as one such constraint. Besides, in the forecasting mode, since it is not straightforward to know a priori the best value of R$_{ss}$, we suggest a path in line with the ensemble approach. Ensemble forecasting is the practice of providing a set of possible future states for the purpose of forecasting a future state. Given the low computational requirement for our modeling framework, which is particularly helpful in ensemble modeling, we suggest for future studies to provide with an ensemble of solar wind profiles at L1 obtained for different R$_{ss}$, with an ensemble of corresponding modeled coronal hole boundaries that are constrained by coronal hole observations. This could be one practical strategy to provide an ensemble of solar wind profiles in an observationally constrained paradigm, which could help us in advancing our solar wind predicting and forecasting capabilities in the future.

    \item We next studied how crucial could be the choice of ADAPT maps for solar wind forecasting, especially since they are produced every two hours. For this, we selected a coronal hole that was observed on 20 August 2015 at the disk center in AIA 193 /AA field of view, which produced an HSS at L1 \citep[see ][]{Reiss2024ApJS}. We found that while a standard GONG CR map produced the modeled boundary of this coronal hole (top panel of Figure~\ref{fig-cr2167-gong}), yet using the ADAPT maps that was created on the same day (20 August 2015), does not produce the modeled coronal hole for any of the 12 ADAPT realizations (Figure~\ref{fig-adapt-1}(a)). However, using the ADAPT maps created 3 days later (23 August 2015) showed that the modeled coronal hole was present in all 12 realization (Figure~\ref{fig-adapt-1}(b)). We also modeled the coronal hole with ADAPT maps created on the following 3 successive days since its appearance at the disk center. We found that this coronal hole was present in the maps created in all three days, however, the sizes of the modeled coronal holes were significantly different on each day (Figure~\ref{fig-adapt-1}(c)). We found that the size of the coronal hole boundary keeps increasing with each day, and it is the third day for which the ADAPT coronal hole boundary is the largest and matches better with the coronal hole boundary obtained from the GONG CR map. Since our results show how sensitive the modeled solar wind profile is to the modeled coronal hole boundaries, it is worth noting that the choice of the ADAPT map (in terms of when they were created) could play an important role in the accuracy of the model output. In this context, we also show the usefulness of coronal hole observations in the context of choosing ADAPT maps, and hope this analysis will work as an important guideline for choosing ADAPT maps for any specific studies in future. We believe this result is not only relevant in the context of real-time solar wind forecasting, but also for models hosted at the Community Coordinated Modeling Center (CCMC) like Enlil and CORHEL which provide ADAPT maps as a choice of input magnetic map for scientific studies. Since ADAPT maps play a crucial role in advancing our current modeling capabilities, improving this particular aspect in the production of future ADAPT maps would be very beneficial. In the future, more detailed studies demonstrating on how to choose the ADAPT maps for a particular study could be extremely useful.
\end{enumerate}

In conclusion, our results indicate that the accuracy of solar wind speed prediction at L1 largely depends on accurately modeling open-closed field line boundaries seen as coronal holes in EUV observations. However, we would also like to point out that this analysis was done using only GONG CR and GONG/ADAPT maps, and the choice of the input magnetic map can also be a significant source of uncertainty as has been pointed out in previous studies (see Section~\ref{intro}). We used a modeling framework similar to those at operational space weather forecasting centers like the Met Office in the UK and SWPC NOAA in the US. Our findings on error sources and uncertainties in ambient solar wind modeling are relevant to both solar wind research and forecasting communities. While our error correction suggestions require further investigation and a quantitative assessment over an extended historical period, we note that an observationally constrained ensemble modeling approach could offer a promising path to more reliable solar wind predictions with reasonable computational demands.

\section{Summary} \label{summary}
This study focuses on exploring the different aspects of the coronal magnetic model that influences key features of an HSS profile at 1 AU. We demonstrated that the grid resolution of the coronal model can strongly influence the starting time ($t_{start}$) and duration ($t_{dur}$) of an HSS at 1 au, and that a better grid resolution leads to better modeling of these two aspects of an HSS. This improvement seems to arise from changes occurring in the modeled coronal hole boundaries for different resolutions which leads to changes in the near-Sun GCAD profile, while the EF profile remains mostly unchanged. We found that changing the source surface height further helped in improving the peak speeds attained during an HSS. In contrast to the case of changing the resolution, a change in R$_{ss}$ changed both the near-Sun GCAD and EF profiles, indicating changes in the coronal hole boundaries and the ambient magnetic morphologies. This was further confirmed by plotting the open field line morphologies for different R$_{ss}$ in the ($r,\phi$) plane. A detailed analysis showed that the improvement due to source surface height optimizations occurs in a two-fold way. Firstly, different source surface heights lead to different sizes (and hence boundaries) of the coronal holes. This change in the coronal hole boundaries lead to different GCAD maps for the coronal holes. Secondly, these differing GCAD values get associated with changing footpoint locations (for different R$_{ss}$) of the open field lines that connect the coronal holes to the sub-Earth points, which leads to differing near-Sun speed map based on the WSA relation. These differing near-Sun speed maps for different source surface heights get essentially propagated to 1 AU through the HUX model and are reflected as different solar wind speed profiles at 1 AU. Thus, our results essentially indicate that for accurate solar wind predictions at L1, accurate modeling of coronal hole boundaries play a crucial role. We further constrain the values of R$_{ss}$ and the corresponding modeled coronal holes with coronal hole observations. By overlaying the footpoint locations on the observed coronal holes, we show that with different R$_{ss}$, we essentially model different regions of the coronal hole that are associated with different GCAD value. We found different values of R$_{ss}$ (ranging between 1.8 - 3.2 R$_{\odot}$) that best models the solar wind speed at L1 for 15 different CRs (that corresponded to different phases of solar cycle 23 and 24). It is also worth noting that predicting a single value for R$_{ss}$ for a particular CR is not straightforward, and also in the forecasting mode it is not easy to know a priori the best R$_{ss}$ value. We thus believe one possible approach in this regard, can be in terms of providing an ensemble of solar wind speed solutions at L1 for different values of R$_{ss}$, where the corresponding ensemble of modeled coronal hole boundaries are observationally constrained with coronal hole observations. We also demonstrate that while working with synchronic maps like ADAPT, the choice of the date of creation of the maps could be crucial in determining the modeling accuracy, and that coronal hole observations can be a useful tool for guiding a better choice of ADAPT maps. 

\section*{Acknowledgements}
The work uses data obtained by the Global Oscillation Network Group (GONG) Program, managed by the National Solar Observatory, which is operated by AURA, Inc., under a cooperative agreement with the National Science Foundation. This work utilizes data produced collaboratively between Air Force Research Laboratory (AFRL) \& the National Solar Observatory (NSO). The ADAPT model development is supported by AFRL. The input data utilized by ADAPT is obtained by NSO/NISP (NSO Integrated Synoptic Program). NSO is operated by the Association of Universities for Research in Astronomy (AURA), Inc., under a cooperative agreement with the National Science Foundation (NSF). The authors thank the SDO team for making the AIA data available and the SOHO team for EIT data. This research was funded in whole or in part by the Austrian Science Fund (FWF) [10.55776/P34437]. For open access purposes, the author has applied a CC BY public copyright license to any author-accepted manuscript version arising from this submission. K.M.~acknowledges support from the NASA cooperative agreement 80NSSC21M0180 and support from the NASA Living With a Star program through its Heliospheric Science Support Office. S.M. acknowledges Yi-Ming Wang for his valuable comments on this study. 



%

\vspace{5mm}
\facilities{NSO/GONG, SOHO/EIT, SDO/AIA, ADAPT, OMNI}


\software{PFSSPY \citep{Stansby2020}
          }



\appendix

\section{Appendix information}


\bibliography{references_sw}{}
\bibliographystyle{aasjournal}



\end{document}